\documentclass[twocolumn]{aastex63}
\usepackage{natbib}
\usepackage{amsmath}
\usepackage{subfigure}
\received{--}
\revised{--}
\accepted{--}
\submitjournal{AJ}

\shorttitle{The LOFAR-MeerKAT-VLA view on the non-thermal side of a jellyfish galaxy}
\shortauthors{A. Ignesti et al.}

\graphicspath{{./}{}}

\begin{document}

\title{GASP XXXVIII: The LOFAR-MeerKAT-VLA view on the non-thermal side of a jellyfish galaxy}

\correspondingauthor{Alessandro ignesti}
\email{alessandro.ignesti@inaf.it}

\author[0000-0003-1581-0092]{Alessandro Ignesti}\affiliation{INAF-Padova Astronomical Observatory, Vicolo dell’Osservatorio 5, I-35122 Padova, Italy}

\author[0000-0003-0980-1499]{Benedetta Vulcani}\affiliation{INAF-Padova Astronomical Observatory, Vicolo dell’Osservatorio 5, I-35122 Padova, Italy}

\author[0000-0001-8751-8360]{Bianca M. Poggianti}\affiliation{INAF-Padova Astronomical Observatory, Vicolo dell’Osservatorio 5, I-35122 Padova, Italy}

\author[0000-0001-9143-6026]{Rosita Paladino}\affiliation{INAF, Istituto di Radioastronomia di Bologna, via Piero Gobetti 101, 40129 Bologna, Italy}

\author[0000-0001-5648-9069]{Timothy Shimwell}\affiliation{ASTRON, the Netherlands Institute for Radio Astronomy, Postbus 2, 7990 AA Dwingeloo, The Netherlands}\affiliation{Leiden Observatory, Leiden University, PO Box 9513, 2300 RA Leiden, The Netherlands}

\author[0000-0003-1020-8684]{Julia Healy}\affiliation{Kapteyn Astronomical Institute, University of Groningen, Landleven 12, 9747 AV Groningen, The Netherlands}\affiliation{Department of Astronomy, University of Cape Town, Private Bag X3, Rondebosch 7701, South Africa}\affiliation{ASTRON, the Netherlands Institute for Radio Astronomy, Postbus 2, 7990 AA Dwingeloo, The Netherlands}

\author[0000-0002-0843-3009]{Myriam Gitti}\affiliation{Dipartimento di Fisica e Astronomia, Università di Bologna, via Piero Gobetti 93/2, 40129 Bologna, Italy}\affiliation{INAF, Istituto di Radioastronomia di Bologna, via Piero Gobetti 101, 40129 Bologna, Italy}

\author[0000-0002-8372-3428]{Cecilia Bacchini}\affiliation{INAF-Padova Astronomical Observatory, Vicolo dell’Osservatorio 5, I-35122 Padova, Italy}

\author[0000-0002-1688-482X]{Alessia Moretti}\affiliation{INAF-Padova Astronomical Observatory, Vicolo dell’Osservatorio 5, I-35122 Padova, Italy}

\author[0000-0002-3585-866X]{Mario Radovich}\affiliation{INAF-Padova Astronomical Observatory, Vicolo dell’Osservatorio 5, I-35122 Padova, Italy}

\author[0000-0002-0587-1660]{Reinout J. van Weeren}\affiliation{Leiden Observatory, Leiden University, PO Box 9513, 2300 RA Leiden, The Netherlands}
\author[0000-0002-0692-0911]{Ian D. Roberts}\affiliation{Leiden Observatory, Leiden University, PO Box 9513, 2300 RA Leiden, The Netherlands}

\author[0000-0002-9325-1567]{Andrea Botteon}\affiliation{Leiden Observatory, Leiden University, PO Box 9513, 2300 RA Leiden, The Netherlands}

\author[0000-0001-9184-7845]{Ancla M\"uller}\affiliation{Ruhr University Bochum, Faculty of Physics and Astronomy, Astronomical Institute, Universit\"atsst 150, 44801 Bochum, Germany}

\author[0000-0003-3255-3139]{Sean McGee}\affiliation{School of Physics and Astronomy, University of Birmingham, Birmingham B15 2TT, United Kingdom}

\author[0000-0002-7042-1965]{Jacopo Fritz}\affiliation{Instituto de Radioastronomía y Astrofísica, UNAM, Campus Morelia, A.P. 3-72, C.P. 58089, Mexico}

\author[0000-0002-8238-9210]{Neven Tomi\v{c}i\'{c}}\affiliation{INAF-Padova Astronomical Observatory, Vicolo dell’Osservatorio 5, I-35122 Padova, Italy}

\author[0000-0002-4382-8081]{Ariel Werle}\affiliation{INAF-Padova Astronomical Observatory, Vicolo dell’Osservatorio 5, I-35122 Padova, Italy}

\author[0000-0003-2589-762X]{Matilde Mingozzi}\affiliation{Space Telescope Science Institute, 3700 San Martin Drive, Baltimore, MD 21218, USA}

\author[0000-0002-7296-9780]{Marco Gullieuszik}\affiliation{INAF-Padova Astronomical Observatory, Vicolo dell’Osservatorio 5, I-35122 Padova, Italy}
\author[0000-0001-9022-8081]{Marc Verheijen}\affiliation{Kapteyn Astronomical Institute, University of Groningen, Landleven 12, 9747 AV Groningen, The Netherlands}



\begin{abstract}
Ram pressure stripping is a crucial evolutionary driver for cluster galaxies. It is thought to be able to accelerate the evolution of their star formation, trigger the activity of their central active galactic nucleus (AGN) and the interplay between the galactic and environmental gas, and eventually dissipate their gas reservoir. We explored the outcomes of ram pressure stripping by studying the non-thermal radio emission of the jellyfish galaxy JW100 in the cluster Abell 2626 ($z=0.055$) by combining LOFAR, MeerKAT, and VLA observations from 0.144 to 5.5 GHz. We studied the integrated spectra of the stellar disk, the stripped tail and the AGN, mapped the spectral index over the galaxy, and constrained the magnetic field intensity to be between 11 and 18 $\mu$G in the disk and $<10$ $\mu$G in the tail. The stellar disk radio emission is dominated by a radiatively old plasma, likely related to an older phase of high star formation rate. This suggests that the star formation was quickly quenched by a factor of 4 in a few $10^7$ yr. The radio emission in the tail is consistent with the stripping scenario, where the radio plasma originally accelerated in the disk is then displaced in the tail. The morphology of the radio and X-ray emissions supports the scenario of accretion of the magnetized environmental plasma onto the galaxy. The AGN non-thermal spectrum indicates that the relativistic electron acceleration may have occurred simultaneously with a central ionized gas outflow, thus suggesting a physical connection between the two processes.

\end{abstract}



\section{Introduction} 

Spiral galaxies residing in clusters  are commonly observed to be redder and having less star formation than field galaxies of similar mass \citep[e.g,][]{Kennicutt1989,Gavazzi2006}, thus indicating that the cluster environment could play the role of `accelerator' of the galaxy evolution \citep[e.g.,][]{Guglielmo_2015}. Environmental processing could be due to either gravitational interactions, in form of tidal interplay with both the cluster and the other galaxies, or hydrodynamical interactions with the environmental plasma \citep[e.g.,][]{Boselli2006,vanGorkom_2004}. The latter, being able to completely remove the gas content of the galaxies, can dramatically impact  their morphology and evolution \citep[e.g.,][for a review]{Boselli_2021}. As galaxies orbit through the cluster, the intracluster medium (ICM) exerts a pressure, known as ram pressure stripping (RPS), on the galactic medium. Such pressure scales as $\rho_\text{ICM}v^2$ \citep[e.g.,][]{Gunn1972} where $\rho_\text{ICM}$ is the mass density of the ICM \citep[typically $10^{-27}$-$10^{-28}$ g cm$^{-3}$ e.g.,][]{Sarazin_1986} and $v$ is the galaxy velocity relative to the ICM  \citep[$\sim500-1000$ km s$^{-1}$  e.g,][]{Cava_2009}. This external pressure can directly strip the interstellar medium (ISM) out of the disk leaving behind a wake of material trailing the galaxy. The gas loss induced by this process can lead to rapid decrease of the star formation rate (SFR) in these galaxies. The most extreme examples of galaxies undergoing strong ram pressure are the so-called jellyfish galaxies \citep[e.g.,][]{Smith2010,Fumagalli2014,Ebeling2014, Poggianti2017}. These objects show extraplanar, one-sided debris visible in the optical/UV light and striking tails of ionised gas. Jellyfish galaxies represent a transitional phase between infalling star-forming spirals and quenched cluster early-type galaxies and provide a unique opportunity to understand the impact of gas removal processes on galaxy evolution. Interestingly, the RPS can also trigger a plethora of (poorly understood) phenomena, such as the formation of extraplanar, cold gas which leads to the formation of new stars outside of the disk \citep[e.g.,][]{Poggianti2019,Gullieuszik_2020}, activity of the central Active Galactic Nucleus (AGN) \citep[e.g.,][Peluso et al., in prep.]{Poggianti2017b}, or complex mixing between ICM and ISM \citep[e.g.,][]{Sun2010,Poggianti_2019,Campitiello_2021}. \\

The radio continuum emission can be a powerful probe to investigate these phenomena. From the radio point of view, cluster late-type galaxies are generally characterized by an excess of radio emission which has been interpreted as a possible evidence of a star formation enhancement due to RPS \citep[e.g.,][]{Gavazzi_1999, Murphy2009,Vollmer_2013,Chen_2020,Roberts_2021a}. In general, radio continuum emission in spiral galaxies is composed of the thermal emission of the $\sim10^4$ K plasma in the HII regions and the non-thermal synchrotron emission of the relativistic cosmic rays electrons (CRe), which are accelerated by supernovae (SN) shocks \citep[e.g.][for a review]{Condon_1992} and reach energies of few GeV. The non-thermal emission provides us with a wealth of information. Being directly related to the SN and due to the fact that galaxies are generally optically thin at GHz frequencies, the non-thermal continuum is a reliable proxy of the SFR \citep[e.g.,][]{Bell_2003,Murphy_2011,Tabatabaei_2017,Gurkan_2018}. Moreover, the ISM microphysics, as well as the magnetic field, can be probed by studying the CRe properties traced by their non-thermal emission \citep[e.g.,][]{Vollmer_2009,Vollmer_2013,Basu_2015,Heesen_2016,Klein_2018,Heesen_2019}. The study of the jellyfish galaxies radio emission recently allowed a series of important advances. The study presented in \citet[][]{Muller_2021} explored  the magnetic field in the tail of the jellyfish galaxy JO206, finding for the first time evidence of an ordered, large-scale field likely induced by the accretion of an envelope of magnetized ICM \citep[this process is also known as magnetic draping, see][for further details] {Dursi_2008,Pfrommer_2010,Sparre_2020}. Moreover, new surveys of radio jellyfish galaxies, made possible also by the advent of the new-generation radio observatory such as the LOw Frequency Array (LOFAR), discovered new populations of galaxies with extended, asymmetrical radio emission both in clusters and groups \citep[][]{Roberts_2021a,Roberts_2021b}. \\

\begin{figure}
    \centering
    \includegraphics[width=\linewidth]{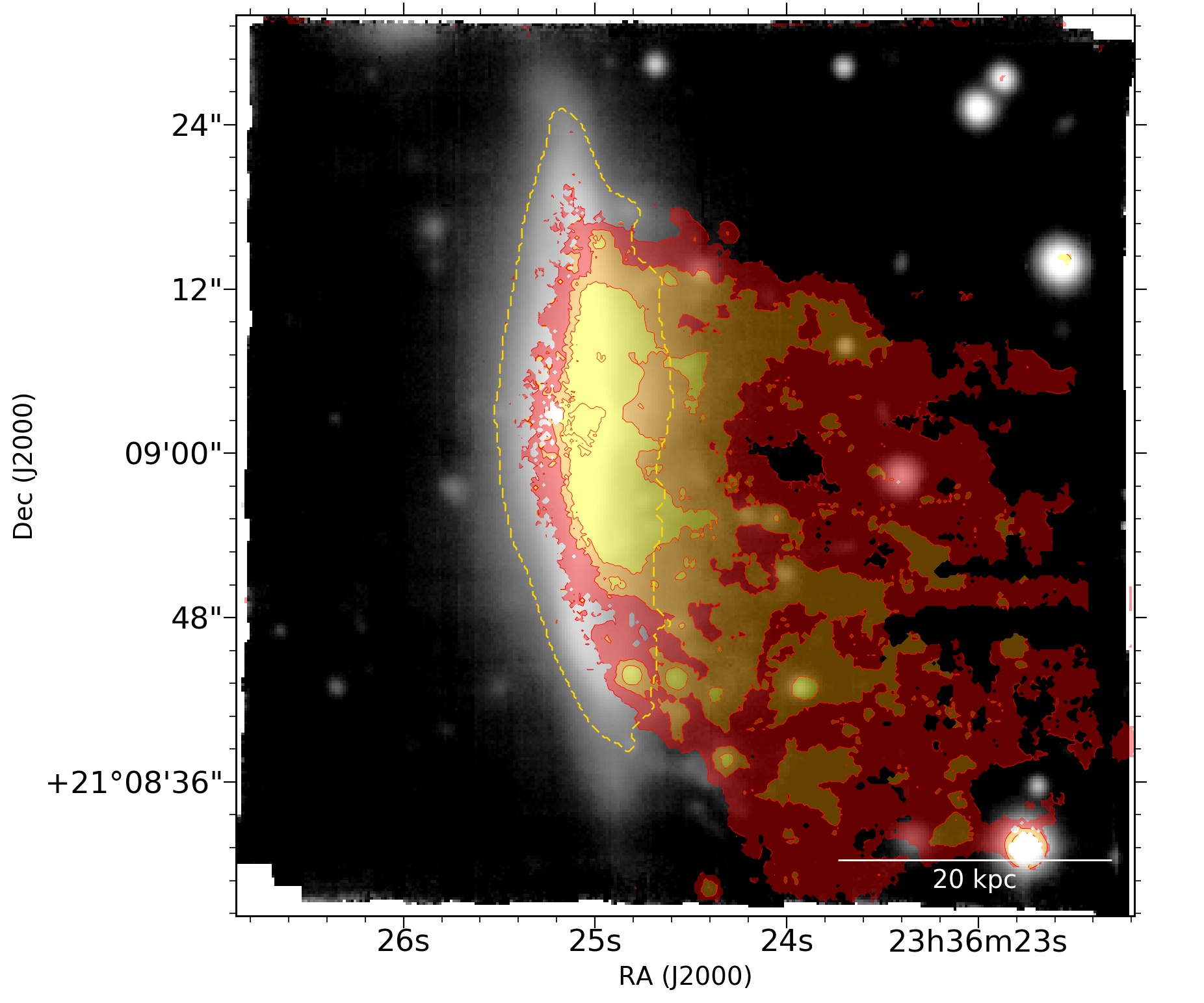}
    \caption{MUSE white light image of JW100 overlaid with} the contours of the H$\alpha$ emission (filled) and the stellar continuum, as defined in \citet[][]{Gullieuszik_2020} (gold dashed).
    \label{opto}
\end{figure}

In this work we explore, for the first time, the non-thermal, multi-frequency radio emission of the jellyfish galaxy JW100 ($z=0.062$, also known as IC5337) in the galaxy cluster Abell 2626 ($z=0.055$), that is a spectacular playground to study the outcomes of RPS. Selected by \citet{Poggianti2016} as a stripping candidate, it is one of the jellyfish galaxies in the GAs Stripping Phenomena \citep[GASP,\footnote{\url{http://web.oapd.inaf.it/gasp/index.html}}][]{Poggianti2017} sample. JW100 is characterized by one of the most striking ionized gas tails in the sample (Figure \ref{opto}), and it is also the most massive galaxy in GASP, with a stellar mass $30 \times  10^{10}$ $M_\odot$ and a total star formation rate (SFR) of $4.0\pm0.8 M_{\odot}\text{ yr}^{-1}$, of which $2.6\pm0.5 M_{\odot}\text{ yr}^{-1}$ is in the tail \citep{Vulcani2018}. Its stellar mass and SFR place JW100 about 0.4 dex below the SFR-mass relation for normal galaxies and $\sim 0.65$ dex below the relation for jellyfish galaxies \citep{Vulcani2018}, indicating that star formation has already decreased due to gas stripping. The effects of RPS in JW100 extends beyond the SFR quenching: in addition to the possible connection between RPS and AGN activity explored in \citet[][]{Poggianti2017b}, \citet[][]{Moretti2018,Moretti_2020} reported the discovery of molecular gas clouds in the tail whose locations are inconsistent with a scenario of stripping from the disk, thus suggesting that they were formed in-situ as a possible consequence of a fast conversion of HI into H$_2$ \citep[see][for further details]{Moretti_2020b}. Finally, by exploring the spatial correlation between the H$\alpha$ and X-ray emission in the tail of JW100, and by studying the spectral properties of the latter, \citet[][]{Poggianti_2019} suggested that the H$\alpha$ emission in the stripped tail \citep[whose optical spectral properties can not be explained by star formation only, see][]{Poggianti2019} could be the result of the complex interplay between ISM and ICM triggered by the RPS.\\

Previous preliminary studies of the radio continuum of JW100, focused on the 0.144 and 1.4 GHz emission, were presented in \citet[][]{Ignesti_2020b} and \citet[][]{Poggianti2019}, respectively. Here we combine multiple radio observations in an unprecedented, multi-frequency study aimed to probe the origin and the properties of the CRe and, by doing so, further explore the history of this galaxy. This paper is structured as follows. In Section 2 we present the datasets and the new images produced for this analysis. In Section 3  the properties of the non-thermal emission inferred from these images are reported, whereas in Section 4 our findings are discussed and interpreted in the context of the galaxy evolution. \\

In this paper we use a \cite{Chabrier2003} initial mass function (IMF) and the standard concordance 
cosmology parameters $H_0 = 70 \, \rm km \, s^{-1} \, Mpc^{-1}$, ${\Omega}_M=0.3$
and ${\Omega}_{\Lambda}=0.7$. At the redshift of Abell 2626 \citep[$z$=0.055, ][]{Cava_2009}, this
yields 1$^{\prime\prime}$=1.071 kpc.

\section{Data analysis}
In this study we present a multi-frequency continuum analysis which combines radio observations at 0.144, 1.4, 3.2 and 5.5 GHz obtained with the LOFAR, MeerKAT and Karl G. Jansky Very Large Array (VLA) radio telescopes, respectively. The data involved in this work have been already published in the literature. We report below a brief description of the data processing of each dataset and the corresponding references, where further details are provided.\\

In order to probe the low-frequency emission at 0.144 GHz, we made use of the LOFAR observation of Abell 2626 presented in \citet[][]{Ignesti_2020b}, which is part of the pointing P353+21 of the LOFAR Two-metre Sky Survey (LoTSS)  \citep[][]{Shimwell_2017,Shimwell_2019}. The dataset was processed using the data-reduction pipeline {\scshape ddf-pipeline} v. 2.2 developed by the LOFAR Surveys Key Science Project\footnote{{\tt https://github.com/mhardcastle/ddf-pipeline}} and an additional self calibration was applied in a smaller region ($\sim1$ deg) centered on the central galaxy IC5338 following the procedure presented in \citet[][]{vanWeeren_2020}. The 1.4 GHz emission was explored by means of a recent MeerKAT observation of the cluster presented in \citet[][]{Healy_2021} (project SCI-20190418-JH-01). The dataset was divided in two bands ($0.960-1.16$ GHz and $1.300-1.520$ GHz) and processed by following the general strategy presented in \citet[][]{Serra_2019} for the cross-calibration using CARAcal \citep[][]{Jozsa_2020}\footnote{\tt https://github.com/caracal-pipeline/caracal}. Here we present the analysis of the $1.300-1.520$ GHz band.\\

Finally, we probed the high-frequency emission by exploiting the VLA observations of the cluster at 3.0 and 5.5 GHz obtained in C-configuration (project code:14B-022, PI Gitti) that are presented in \citet[][]{Ignesti_2017}. The data reduction was done using the National Radio Astronomy Observatory (NRAO) Common Astronomy Software Applications package \citep[CASA, ][]{McMullin_2007}, version 4.6. Due to  flagging of radio-frequency interference, the bandwidth of the 3 GHz observation decreased from 2.0 GHz ($2.0-4.0$ GHz) to 1.6 GHz ($2.4-4.0$ GHz), thus moving the central frequency from 3.0 GHz to 3.2 GHz. \\

\begin{figure*}
    \centering
\includegraphics[width=\linewidth]{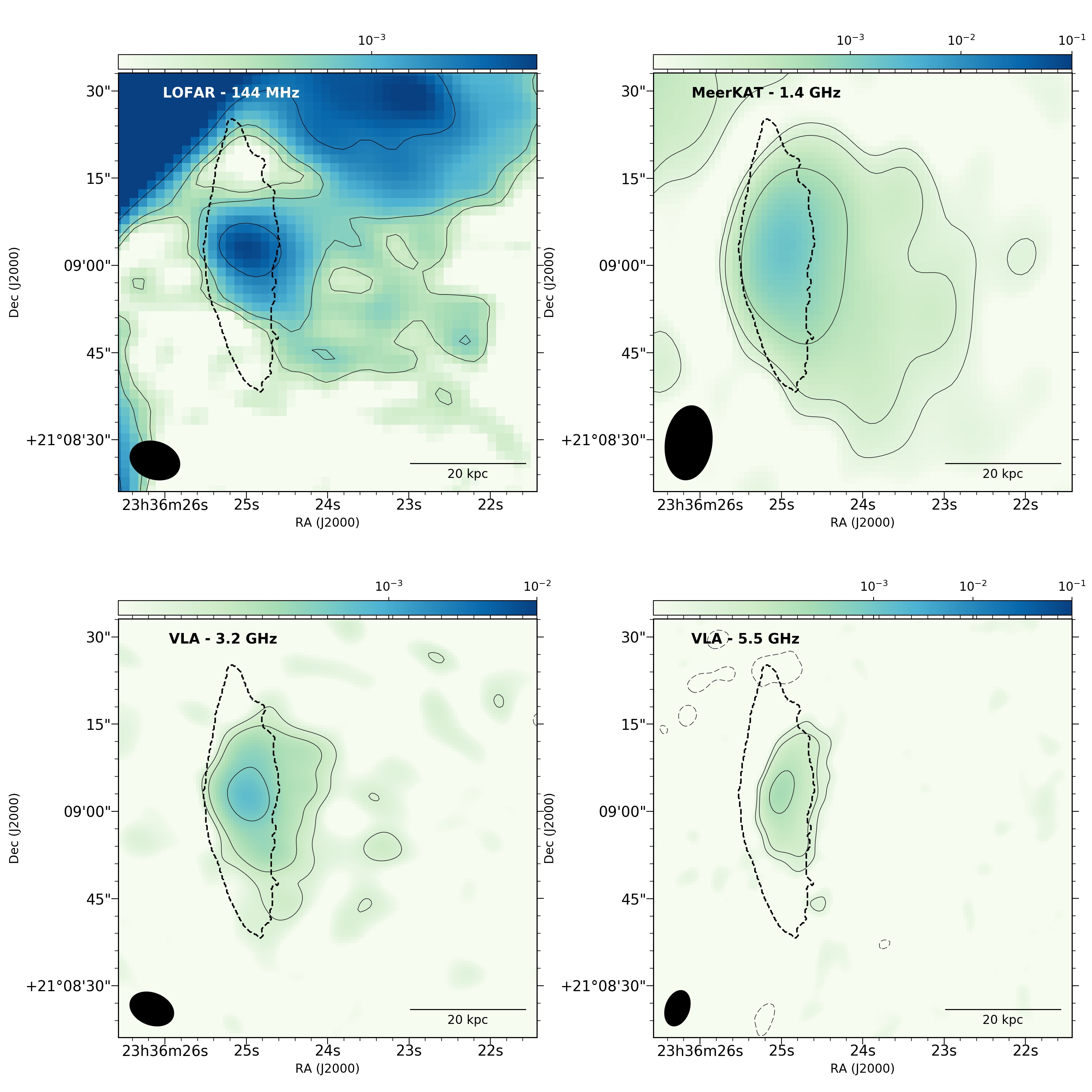}
    \caption{\label{canvas}Images of JW100 at different frequencies with overlaid the contours corresponding to the -3, 3, 6, 12, 24$\times$RMS levels (continuous) and stellar continuum emission (dashed). The respective RMS and resolution (shown by the black filled ellipses at the bottom-left corner of each image) at each frequency are reported in Table \ref{prop}}.
   
\end{figure*}

\begin{figure*}
    \centering
\includegraphics[width=\linewidth]{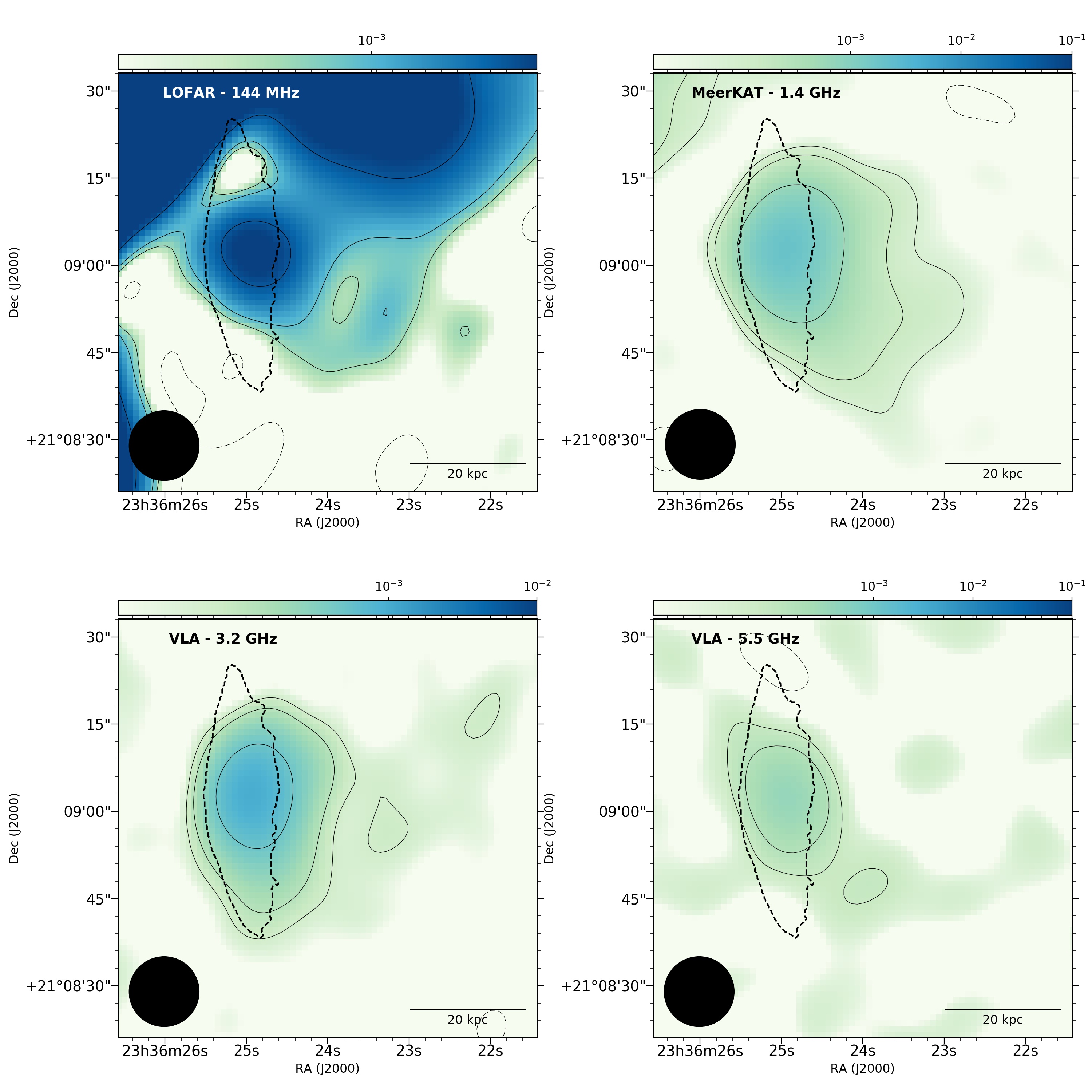}
    \caption{\label{canvas_smooth}Images of JW100 at different frequencies smoothed with a 12$''\times$12$''$ Gaussian beam which is shown in the bottom-left corner of the images.  We report also, on top, the contours corresponding to the -3, 3, 6, 12, 24$\times$RMS levels (continuous) and stellar continuum emission (dashed). The respective RMS at each frequency are reported in Table \ref{prop}.}
    
\end{figure*}

We produced new images of JW100 at different frequencies using {\scshape WSClean} v2.10.1 \citep[][]{Offringa_2014}. We tested different combinations of Briggs weightings \citep[][]{Briggs_1994}, with {\ttfamily robust} values ranging from 0 to -2, and multi-scale cleaning \citep[][]{Offringa_2017}. For the LOFAR data we adopted a lower UV-cut of 80$\lambda$, corresponding to an angular scale of 43$^\prime$,  to drop the shortest spacings where calibration is more challenging \citep[see][for further details about the imaging of LoTSS observations]{Shimwell_2017}. The final images are presented in  Figure \ref{canvas}, whereas  in Table \ref{prop} we report the corresponding resolution and root mean square (RMS)  noise. The latter was measured for each map by using the same set of rectangular boxes manually placed on regions of the sky within 15 arcmin from JW100 and devoid of other radio sources. Moreover, in order to reliably compare the flux density at different frequencies, we produced a second set of images with matched UV-range ($670-19000$ $\lambda$) and resolution ($12''\times12''$), which we report in  Figure \ref{canvas_smooth}. In order to reliably compare the signal collected by the different UV-sampling configurations, we adopted an {\tt UNIFORM} weighting for MeerKAT and VLA data. For the LOFAR data a different weighting was adopted
({\tt ROBUST=-1}), to detect the extended emission, which  was lost using  the UNIFORM weight \citep[see ][Figure 1,  bottom-left panel]{Ignesti_2020b}. For consistency, the RMS levels of these images (Table \ref{prop}) were evaluated by using the same set of regions adopted for the high-resolution images.\\

We jointly analyzed the radio emission of JW100 with a Multi Unit Spectroscopic Explorer (MUSE) optical observation to carry out a multi-wavelength analysis of the galaxy. The MUSE observations, data reduction and the methods of analysis are described in \citet{Poggianti2017}. Specifically, we analyzed the nebular Balmer emission line of $\rm H\alpha$, which is measured from the MUSE datacube with both the optically-thin corrections for the Galactic foreground dust extinction, and the intrinsic dust extinction and stellar absorption. Finally, we included also the high-resolution VLA observation at 1.4 GHz presented in \citet[][]{Gitti_2013b} (1.3 $''$) to study the morphology of the AGN radio emission, and the X-ray {\it Chandra} observation of Abell 2626 \citep[][]{Ignesti_2018} to evaluate the spatial correlation between the distribution of the relativistic and hot plasmas.

\begin{table}[]
    \centering
   
    \begin{tabular}{ccc|c}
    \hline
    Frequency&Resolution& RMS$^{1}$ & RMS$^{2}$\\
    $[$GHz$]$&[arcsec]&[$\mu$Jy beam$^{-1}$ ]&[$\mu$Jy beam$^{-1}$]\\
    \hline
    0.144&8.8$\times$ 6.4 &93.8&142.2\\
    1.4&12.4$\times$ 7.7 &16.2&24.4\\
    3.2&7.8$\times$ 5.4& 15.0&20.4\\
    5.5&6.3$\times$ 4.1& 6.0&25.4\\
    \hline
    \end{tabular}
     \caption{Properties of the radio images reported in Figure \ref{canvas} and \ref{canvas_smooth}. (1) and (2) indicate the RMS measured before  and after the smoothing to the 12$''\times$12$''$ resolution.}
    \label{prop}
\end{table}

\section{Results}
\subsection{Radio continuum images}
The images reported in Figures \ref{canvas} and \ref{canvas_smooth} show that the morphology of the emission changes with frequency. The radio emission extends from the stellar disk toward the south-west, roughly following the morphology of the H$\alpha$ emission (Figure \ref{opto}). At the current resolution and sensitivity, the extra-planar radio emission, i.e. the radio tail, projected length (here defined as the extension of the radio emission from the edge of the stellar disk up to the farthest 3$\times$RMS level contour outside the disk, along the direction of the H$\alpha$ tail, measured in the high-resolution images and not deconvolved for the beam smearing) decreases from $\sim$37 kpc at 144 MHz to almost zero at 5.5 GHz. Whilst the disk emission is clearly distorted at 3.2 and 5.5 GHz, the tail is too faint to be confirmed at these frequencies. Indeed by assuming a spectral index $\alpha=-0.7$ and given the current RMS levels at 1.4, 3.2, and 5.5 GHz (Table \ref{prop}), the $3\times$RMS level contours of the 0.144 GHz image would  correspond to detection levels of 3.5$\times$RMS, 2.4$\times$RMS, and 1.3$\times$RMS, respectively. Absence of a clear detection of the tail at the 3.2 and 5.5 GHz is therefore not a surprise.\\

Within the stellar disk, the continuum emission is more extended in the north-south direction at 1.4, 3.2 GHz than at 0.144 and 5.5 GHz. In contrast, two features, i.e. the presence of the AGN radio emission at the center and the symmetrically truncated emission with respect to the stellar disk, can be observed at every frequency. Examples of radio disk truncation have already been reported in the literature \citep[e.g.,][]{Vollmer_2009,Vollmer_2013,Chen_2020,Roberts_2021a} and they have been claimed to be a result of ram pressure stripping. The implications of the morphology of the radio emission are discussed in Section \ref{morph}.\\

Finally, we note that the radio emission in the top-left corners of the LOFAR and MeerKAT images is part of the Kite radio source located at the center of Abell 2626 \citep[][and references therein]{Ignesti_2020b}. Albeit we do not observe significant negative artifacts around the galaxy in the 0.144 GHz image (which indicates that the cleaning of the brighter Kite source did non affect dramatically the morphology of JW100), the resolution of our image merges the two sources, both to the east of the galaxy and the northern part of the tail. As we discuss in the following Section, we carefully accounted for this during the study of the radio tail.
\subsection{Resolved radio properties}
\label{resolved}
\begin{figure}
    \centering
    \includegraphics[width=\linewidth]{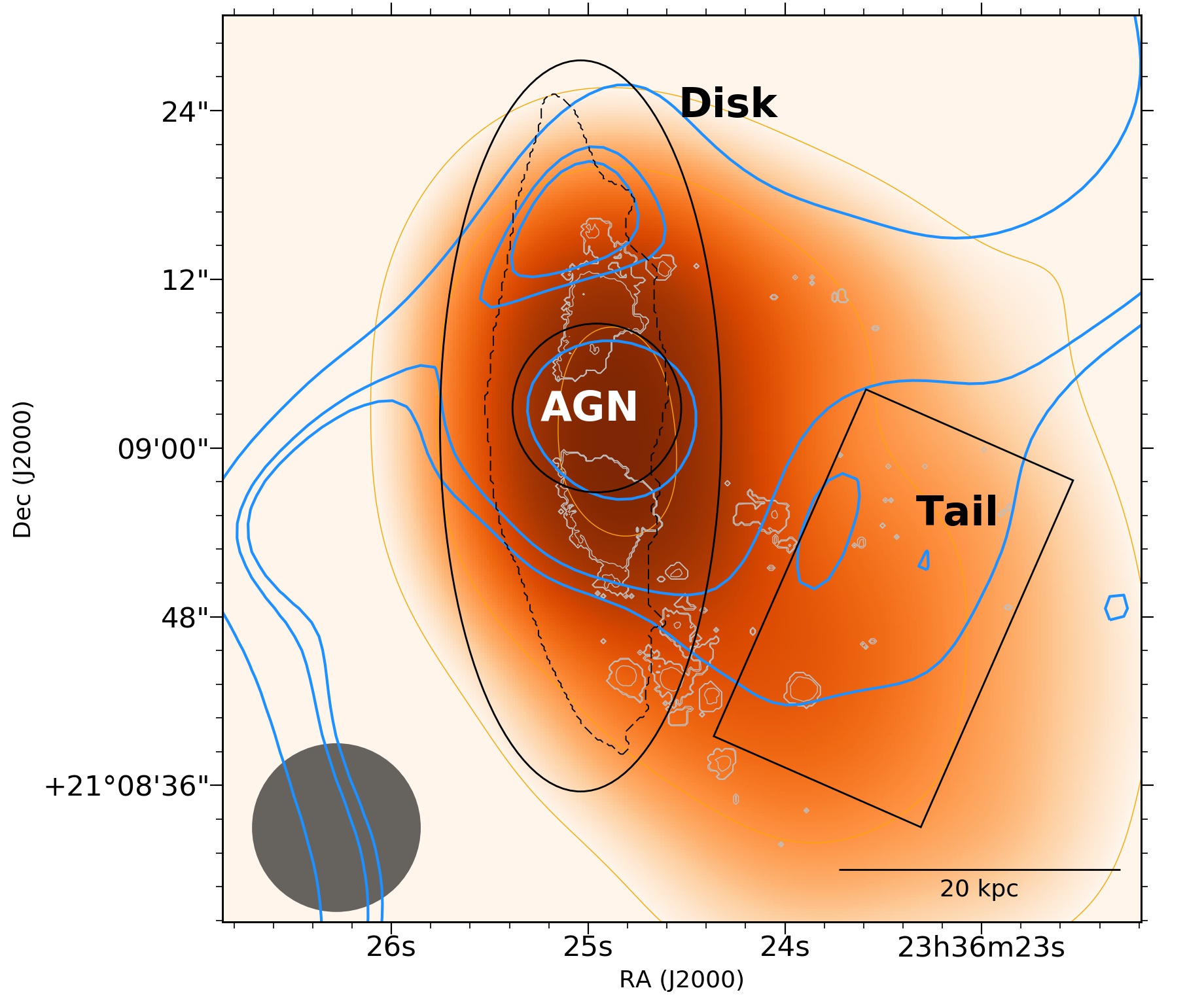}
    \caption{MUSE H$\alpha$ emission (Figure \ref{opto}) smoothed to 12$''$ (color image and orange contours) with the contours of stellar continuum (black dashed), the spaxels classified as star forming \citep[silver contours, see][]{Poggianti2019}, and the 3 regions (solid black geometrical shapes). For reference, we report also the positive contours of the 0.144 GHz image shown in Figure \ref{canvas_smooth} (blue)  and the beam shape (black-filled circle in the bottom-left corner)}.
    \label{halpha}
\end{figure}

The resolution reached in these observations allows us to explore the properties of the radio emission in 3 different regions, that are the AGN, the stellar disk, and the stripped tail. These regions, which are reported in Figure \ref{halpha}, were defined as follows:
\begin{itemize}
    \item AGN: we defined a $12''$ ($\sim 12\times 12$ kpc) diameter circular region centered on the surface brightness peak observed at 0.144 MHz that coincides with the point source studied in \citet[][]{Poggianti2017b, Radovich2019, Poggianti_2019}. Its properties are discussed in Section \ref{agn};
    \item Stellar disk: \citet[][]{Gullieuszik_2020} defined the stellar disk of JW100 on the basis of the H$\alpha$ emission observed with MUSE. Nonetheless, due to the difference in resolution between optical and radio images, we have to define the size of the stellar disk accordingly. We produced a smoothed H$\alpha$ image to match the resolution of the smoothed radio images, then we estimated the size of the smoothed stellar disk by defining an ellipse that contains the same H$\alpha$ flux measured within the stellar disk in the original image. This allows us, for each radio map, to define the disk emission as the radio flux density above the 3$\times$RMS level measured within the elliptical region shown in Figure \ref{halpha}. For the 0.144 GHz image we did not include the emission of the Kite present in the northern part of the elliptical region;
    \item Tail: we defined a region along the H$\alpha$ tail that is far enough to minimize the possible contribution from the disk and, at 0.144 GHz, from the northern plume of the Kite radio source. At 0.144 and 1.4 GHz we clearly resolve extended radio emission within this region, whereas at 3.2 GHz we have only a tentative detection, thus we estimated a reference value for the flux density by multiplying  the 3$\times$RMS level for the area of tail region. This provided us a generous upper-limit to prudently constrain the non-thermal spectrum up to 3.2 GHz. At 5.5 GHz, in the low-resolution image only, we observe a component of emission in this region, with an angular size below the resolution of the map. Due to the fact that it does not spatially coincide with the emission observed at 3.2 GHz, we can not exclude that it is an artifact, and hence we conservatively decided to restrict the analysis of the stripped tail to the 0.144-3.2 GHz band.
\end{itemize}
These regions were used to measure the radio flux densities in the smoothed radio images (Figure \ref{canvas_smooth}). We assumed calibration errors of 20, 10, 5 and 5$\%$ for the emission at 0.144, 1.4, 3.2 and 5.5 GHz, respectively.  We computed the net flux density in the disk by subtracting, at each frequency, the AGN contribution from the total emission. Therefore, in the following we refer to `disk emission' as the net radio emission within the stellar disk devoid of the AGN contribute. \\

As mentioned in Section 1, the total radio emission is the result of the combination of the non-thermal CRe synchrotron emission and the thermal radio emission provided by the warm ISM. Therefore, to estimate the non-thermal emission, $S_{nt}$, we have to subtract the optically-thin thermal component, $S_{th}$, to the total measured flux densities, $S_{tot}$:  $S_{nt}=S_{tot}-S_{th}$. The thermal flux, $S_{th}$, can be expressed as:
\begin{equation}
    S_{th}=1.14\times10^{-14} \left(\frac{T_e}{10^4\text{ K}}\right)^{0.34} \left(\frac{\nu}{\text{GHz}}\right)^{-0.1}S_{\text{H}\alpha}
\end{equation}
where $S_{\text{H}\alpha}$ is the H$\alpha$ flux measured in the smoothed H$\alpha$ image (Figure \ref{halpha}) in units of erg s$^{-1}$ cm$^{-2}$, and $T_e$ is the thermal electron temperature, which we assumed to be $10^4$ K \citep[e.g.,][]{Deeg_1997,Tabatabaei_2017}. We note that, due to the complex interplay between ISM and ICM which is taking place in this galaxy, the ionized ISM could potentially reach temperatures up $10^6$ K \citep[e.g.,][]{Kanjilal_2020}, hence the thermal radio emission might be more relevant. The current data does not allow us to reliably determine the temperature of the ionized ISM, hence we can not exclude that we may be underestimating the contribution of the thermal radio emission, especially in the tail where the ISM-ICM interplay is potentially more significant \citep[see][]{Poggianti_2019}. 

\begin{figure}
\centering
         \includegraphics[width=\linewidth]{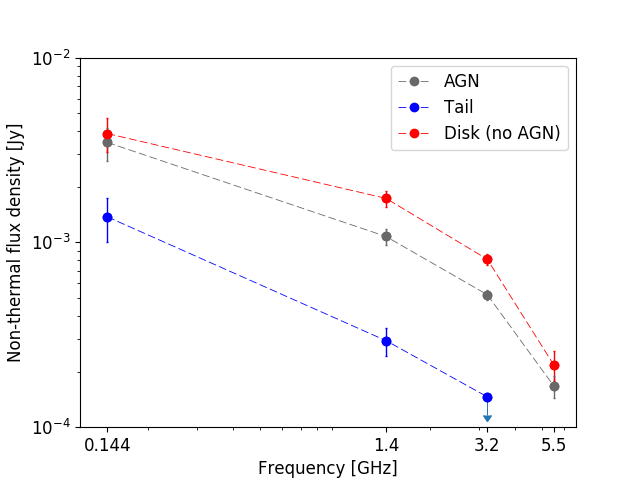}
         \includegraphics[width=\linewidth]{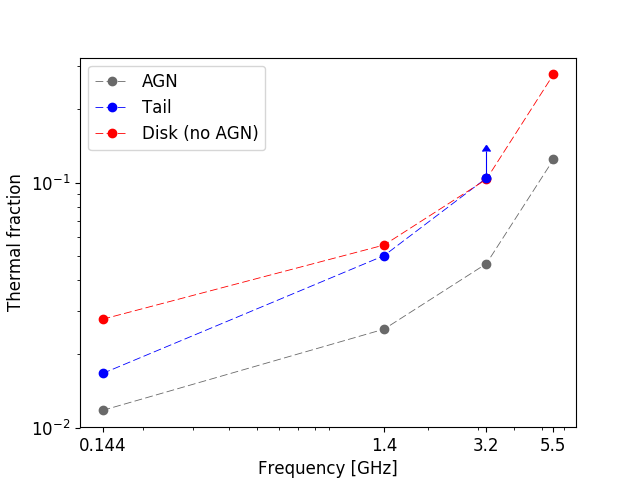} 
    \caption{Non-thermal spectra (top) and thermal fractions (bottom) for AGN, disk and stripped tail. At 3.2 GHz  we report the upper limits of the stripped tail non-thermal flux density and thermal fraction. }
    \label{spectra}
\end{figure}

\begin{table}
\centering
  \begingroup
\setlength{\tabcolsep}{4pt}
    \begin{tabular}{ccccc}
    
    \multicolumn{5}{c}{\bf Non-thermal flux density $S_{nt}$ [mJy]}\\
    \hline
    Region&0.144 GHz&1.4 GHz&3.2 GHz&5.5 GHz\\
    \hline
    AGN &3.48$\pm$0.71& 1.00 $\pm$0.11& 0.52$\pm$ 0.03& 0.17$\pm$ 0.02\\
    Disk&3.88$\pm$ 0.81& 1.73$\pm$ 0.18& 0.81$\pm$ 0.06& 0.22$\pm$ 0.04\\
    Tail& 1.37$\pm$ 0.37 &0.29$\pm$ 0.05& $<$0.15&$\#$\\
    \hline
    &&&&\\
    \multicolumn{5}{c}{\bf Thermal fraction $f_{th}$}\\
    \hline
    Region&0.144 GHz&1.4 GHz&3.2 GHz&5.5 GHz\\
    \hline
    AGN&0.01&0.03& 0.05& 0.12\\
    Disk&0.03& 0.06& 0.10& 0.28\\
    Tail&0.02&0.05& $>$0.10&$\#$\\
    \hline
    &&&&\\
    \multicolumn{5}{c}{\bf Non-thermal spectral index $\alpha_{\nu_1}^{\nu_2}$}\\
    \hline
    Region&$0.144-1.4$&\multicolumn{2}{c}{$1.4-3.2$}&$3.2-5.5$\\
    \hline
    AGN&-0.52$\pm$ 0.10&\multicolumn{2}{c}{ -0.88$\pm$ 0.14}& -2.10$\pm$ 0.28\\
    Disk&-0.35$\pm$ 0.10&\multicolumn{2}{c}{ -0.92$\pm$ 0.15}& -2.43$\pm$ 0.37\\
    Tail&-0.68$\pm$ 0.14&\multicolumn{2}{c}{ $<-0.85$}&$\#$\\
    \hline
    \end{tabular}
    \endgroup
    \caption{Results of the resolved analysis. For each region we report the non-thermal radio flux densities (mJy, top), the thermal fraction (middle), and the non-thermal spectral indexes (bottom). Those values are not reported for the tail at 5.5 GHz due to the concerns on the reliability of the detection.   }
    \label{result_1}
\end{table}
In Figure \ref{spectra} and Table \ref{result_1} we report the resulting non-thermal flux densities, the fractions of the thermal radio flux with respect of the total ($f_{th}=S_{th}/S_{tot}$), and the non-thermal spectral index in the 3 bands ($0.144-1.4$, $1.4-3.2$, and $3.2-5.5$ GHz) computed as:
\begin{equation}
    \alpha_{\nu_1}^{\nu_2}=\frac{\text{log}\left(\frac{S_1}{S_2}\right)}{\text{log}\left(\frac{\nu_1}{\nu_2}\right)}\pm \frac{1}{\text{log}\left(\frac{\nu_1}{\nu_2}\right)}\sqrt{\left(\frac{\sigma_1}{S_1}\right)^2+\left(\frac{\sigma_2}{S_2}\right)^2}
    \label{spec}
\end{equation}
where $\nu$, $S$ and $\sigma$ are the frequency, the  non-thermal flux density and the corresponding error, respectively. \\

The non-thermal integrated spectra of the three components denote a series of interesting features:
\begin{itemize}
 \item The AGN spectrum steepens from $\alpha=-0.52\pm 0.10$ at low frequencies to $\alpha=-2.10\pm 0.28$ at 5.5 GHz. As a caveat, we note that the combination of low angular resolution and edge-on view of the disk does not allow us to completely rule out a contamination from the disk emission. We discuss the AGN non-thermal spectrum in the context of the previous studies in Section \ref{agn};
\item The synchotron emission in the disk shows both an evident flattening at low frequencies ($\alpha=-0.35\pm 0.10$) and a strong steepening at higher frequencies. The low frequency flattening is expected as consequence of the ionization losses of low-energy CRe in the high-density, star-forming regions \citep[e.g.][]{Murphy_2009b, Basu_2015,Chyzy_2018}. The steepening at high frequencies indicates that the relativistic plasma is old enough to have been affected by significant energy losses \citep[e.g.,][]{Pacholczyk_1970, Klein_2018}.  Interestingly,  \citet[][]{Vulcani2018,Poggianti2019} report on the ongoing star formation within the disk, that would entail the injection of fresh relativistic plasma emitting with a spectral index $-0.7<\alpha<-0.5$ up to 5.5 GHz. Therefore, the steep integrated spectrum suggests an inhomogeneity in the CRe distribution, so that the steep-spectrum emission results due to the mixing of old and fresh CRe within the disk \citep[e.g.,][]{Chyzy_2018,Heesen_2019}. The connection between non-thermal radio emission and SFR is further discussed in Section \ref{sfr_disc};
\item The spectrum of the tail is steeper than the disk in the 0.144-1.4 GHz band  ($-0.68\pm0.14$ vs $-0.35\pm0.10$), and potentially also in the 1.4-3.2 GHz one ($<-0.85$ and $-0.92\pm0.15$). This is in agreement to what is reported in \citet[][]{Muller_2021} for the jellyfish galaxy JO206.
\end{itemize}
Finally, the thermal fractions increase with frequency, in agreement with the results reported in literature \citep[e.g.,][]{Tabatabaei_2017}.  

\subsection{Spectral index maps}
\label{spectral_mapss}
\begin{figure*}
\centering
{\bf 0.144$-$1.4 GHz} \hfill\\
\vspace{0.3 cm}
    \includegraphics[width=.46\linewidth]{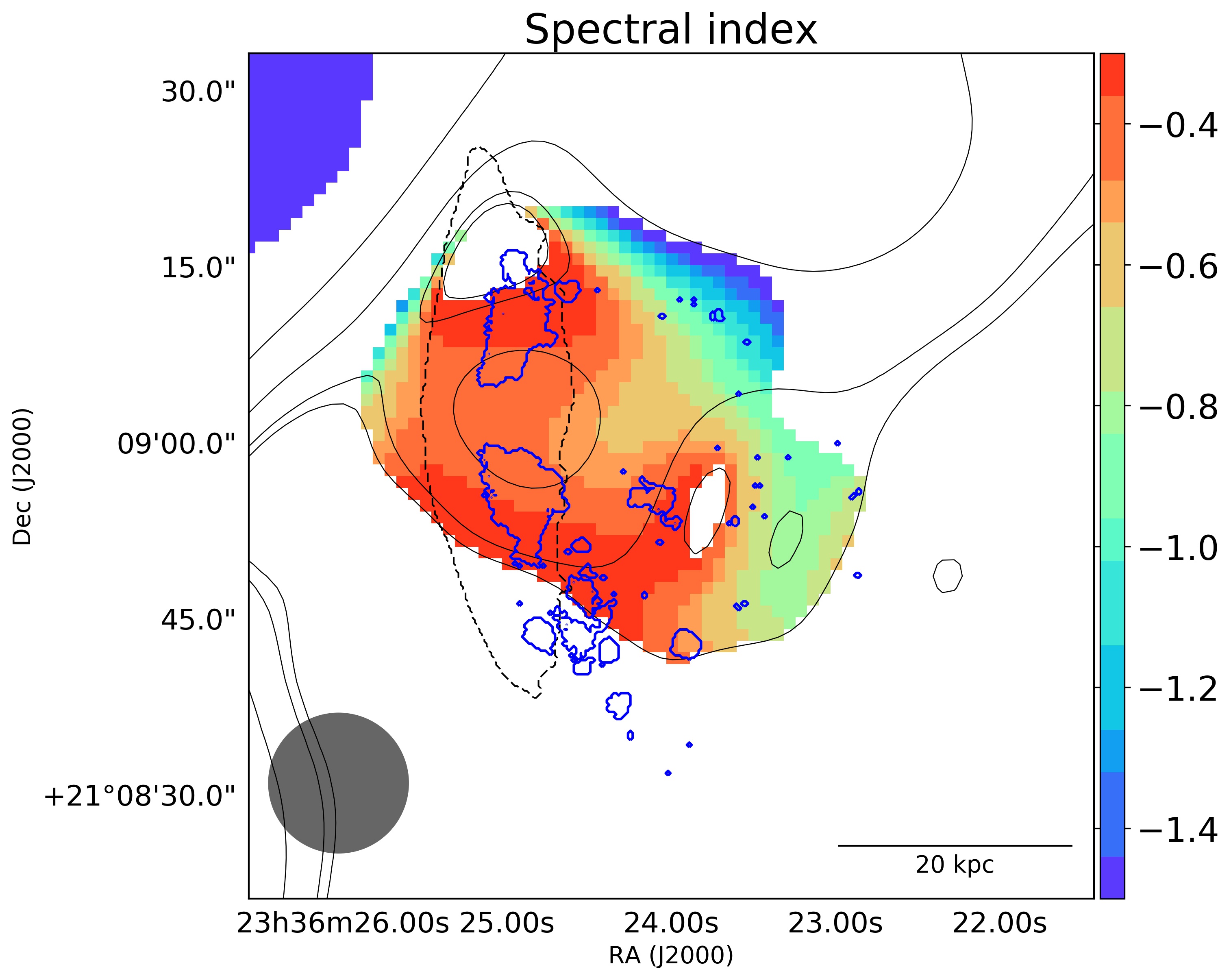}
    \includegraphics[width=.46\linewidth]{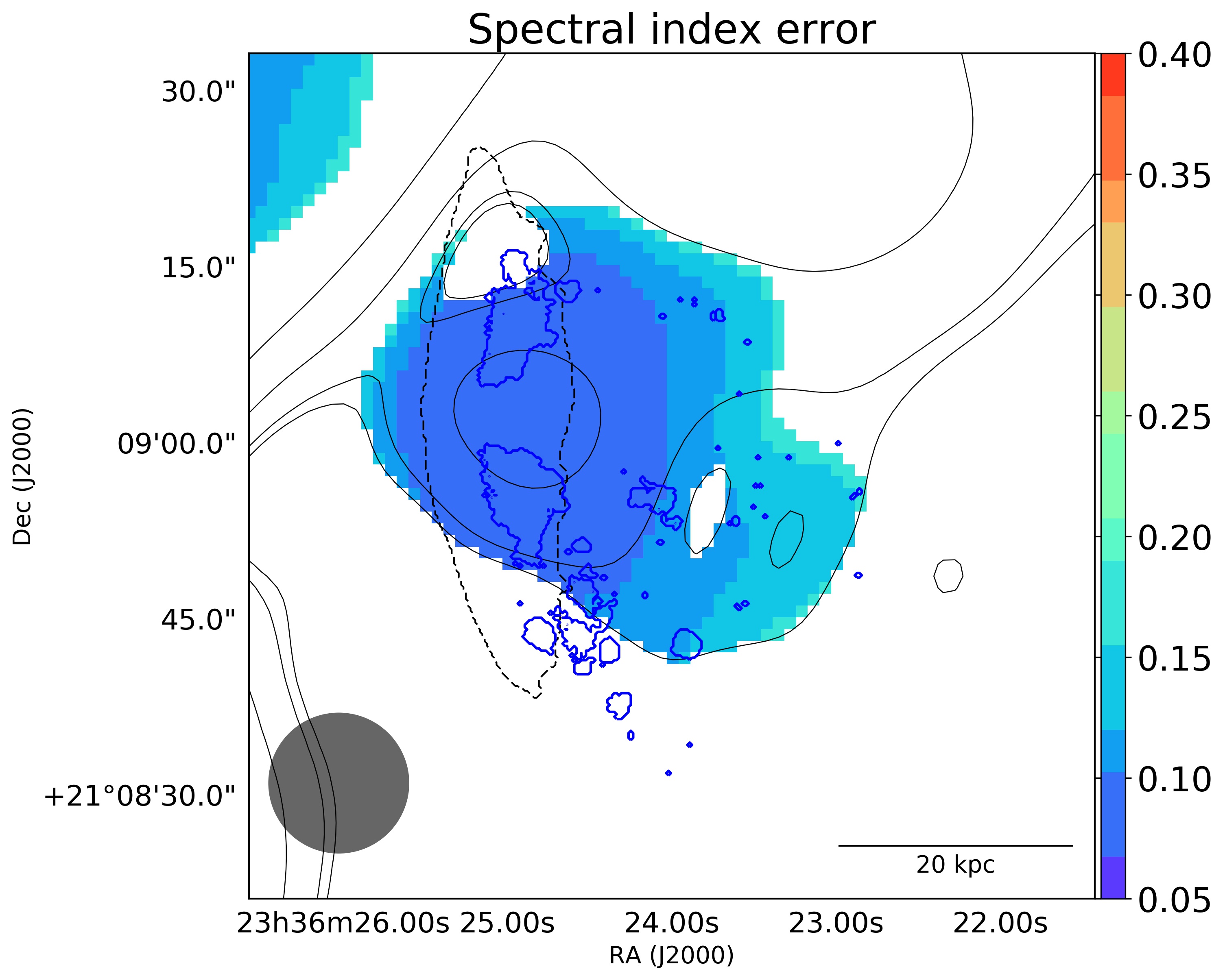}
\\ {\bf 1.4$-$3.2 GHz} \hfill\\
\vspace{0.3 cm}
    \includegraphics[width=.46\linewidth]{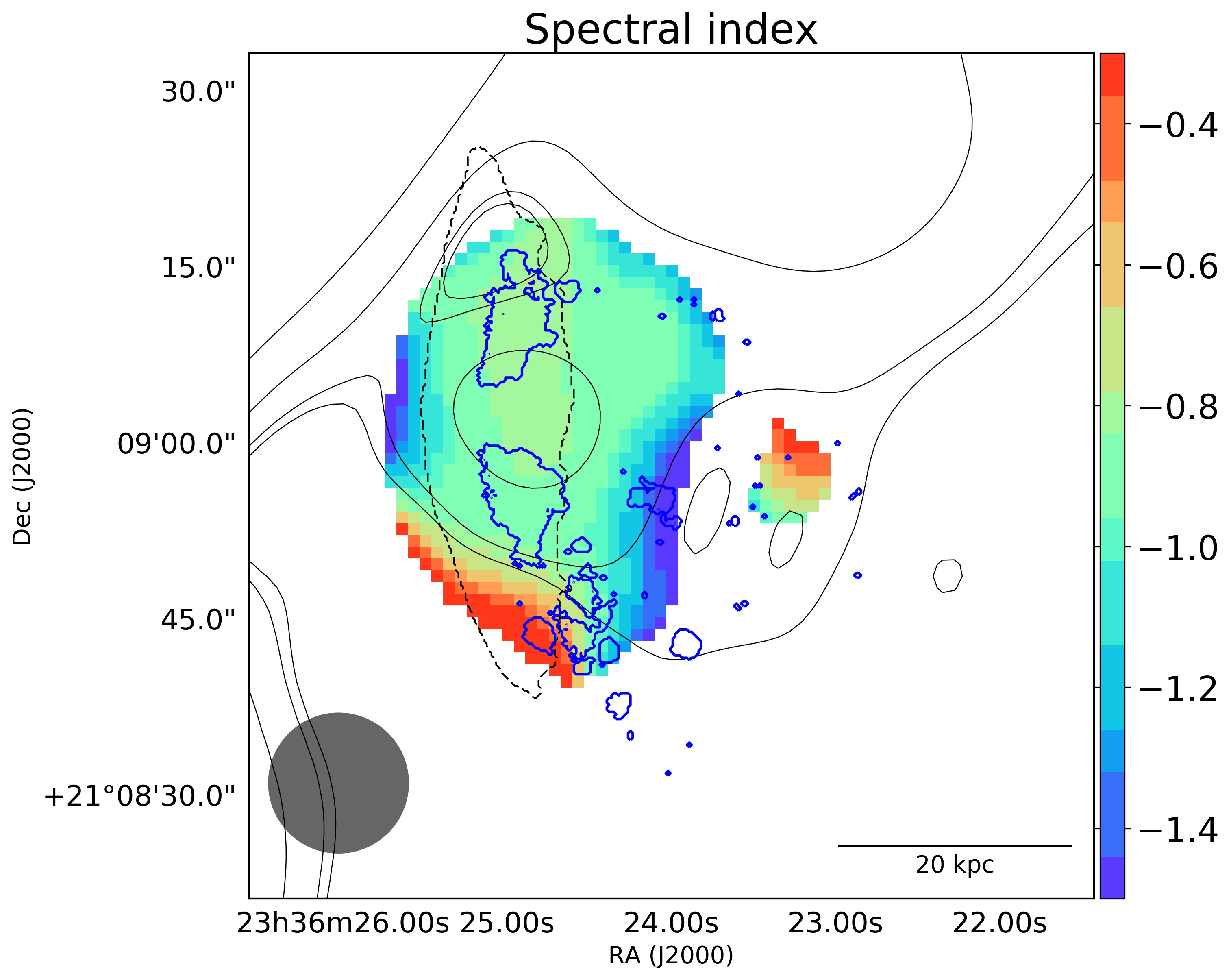}
    \includegraphics[width=.46\linewidth]{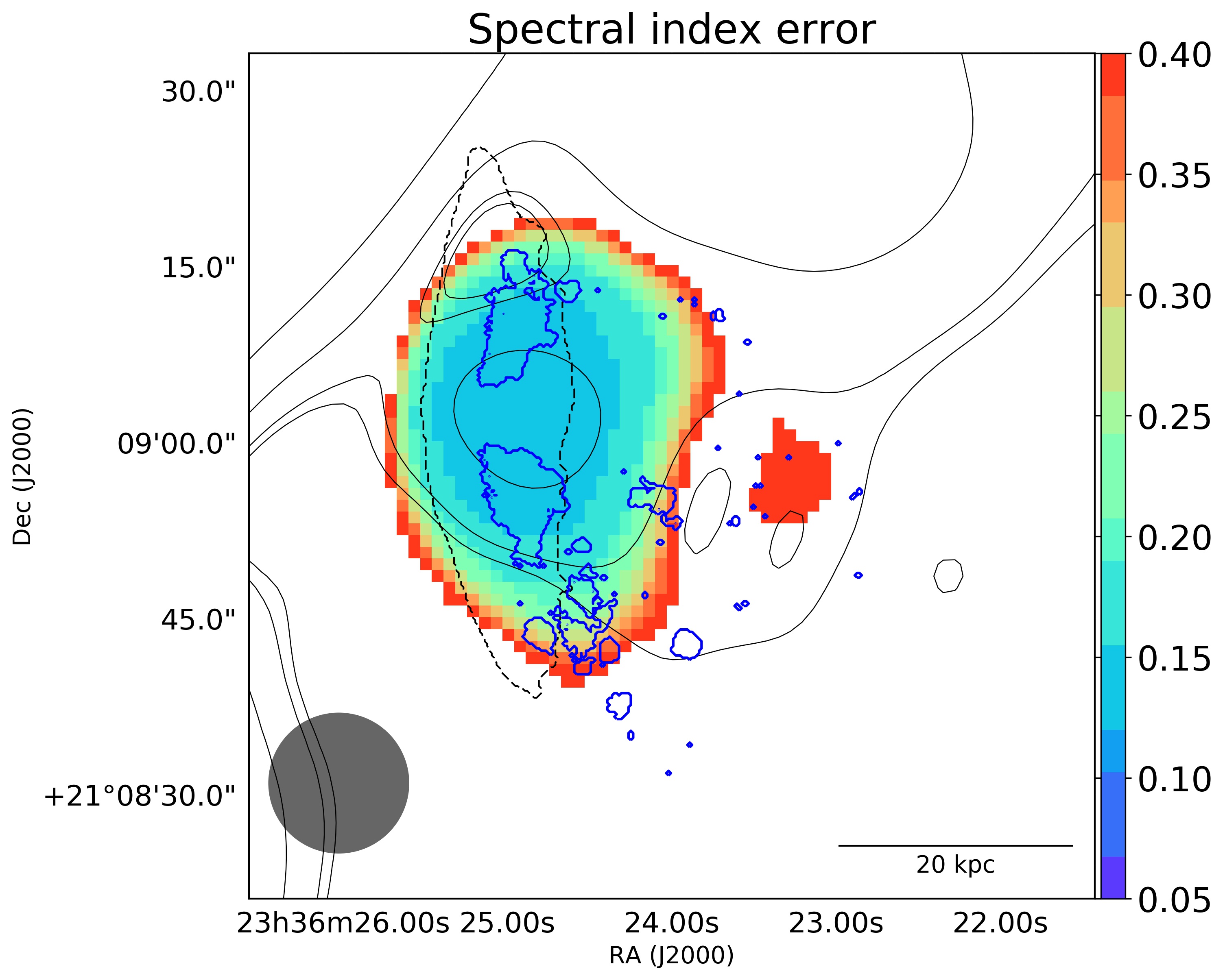}

    \caption{\label{spectra_map}Spectral index (left) and error (right) maps for 0.144-1.4 GHz (top) and 1.4-3.2 GHz (bottom) with a resolution of 12$''\times$12$''$ (bottom-left corner). In each map we report also the contours of the 144 MHz emission (black continuous), stellar continuum (black dashed), and the contours of the star forming regions reported in Figure \ref{halpha} (blue).}
    
\end{figure*}

We combined the smoothed images at 0.144, 1.4 and 3.2 GHz (Figure \ref{canvas_smooth}) to map the spectral index total radio emission from the disk to the tail in the 0.144-1.4 and 1.4-3.2 GHz bands. We excluded the 5.5 GHz map from this analysis because it does not provide significant information on the extended emission. For each map, we selected the emission above the respective 3$\times$RMS levels and we computed the spectral index pixel-by-pixel following Equation \ref{spec}. The thermal fractions reported in Figure \ref{spectra} show that the thermal contribution below 3.2 GHz is negligible, thus the resulting values are well representative of the non-thermal component. The resulting images and the corresponding error maps are reported in Figure \ref{spectra_map}.\\

As seen from the flux density measurements, the spectral index maps show a generally flat spectrum in the disk ($\alpha>-0.7$), and a steepening towards the tail ($\alpha<-0.8$). In the 0.144-1.4 GHz maps we observe that the flat-spectrum regions roughly coincide with the star forming regions selected according to the [O I] BPT diagrams reported in \citet[][]{Poggianti2019} (see Figure \ref{halpha}) \footnote{We choose the [O I] over the [S II] diagram to be more conservative because it disentangles the star forming regions in the tail from the rest of the H$\alpha$ emission, whose origin is unclear \citep[see][for a detailed discussion]{Poggianti_2019}.}, both within and outside of the stellar disk, hence suggesting that there the CRe are younger, i.e. freshly injected. The flattest spectrum regions ($\alpha\simeq -0.3\mp0.1$) are located at the disk edges, in agreement withe the integrated spectrum (Table \ref{result_1}). The spectral steepening trend outside the disk resembles the previous results presented in literature \citep[e.g.,][]{Vollmer_2004,Muller_2021} and it can be evidence that the plasma in the tail is generally older than in the disk, i.e. that the CRe are accelerated in the disk and then stripped away along the tail. We note that the steepest spectrum ($\alpha\simeq$-1.4) is observed in the direction of the northern, ultra steep-spectrum plume of the Kite radio source \citep[see][]{Ignesti_2020b}. On the basis of the striking difference in the spectral indexes between the galaxy and the plume (-1.4 vs. $<$-3) we argue that the steepening is likely due to the combination of low-resolution and projection effects which added up  the relatively-flat spectrum tail with the ultra-steep emission of the Kite. The  1.4-3.2 GHz spectral map is generally more uniform, with values between -1.0 and -0.8. The southern edge of the disk exhibits a flat-spectrum edge ($\alpha\simeq-0.4\pm0.4$) which is barely in agreement with the integrated spectrum ($\alpha=-0.92\pm0.15$), thus we suggest that it may be an artifact produced by the similar RMS level of the two images.

\subsection{Magnetic field}
\label{magnetic}
Observing non-thermal radio emission implies that the galaxy is permeated by a magnetic field. Furthermore, the asymmetrical morphology points out that the CRe within the tail are moving within a magnetic field that may either be extending from the disk, or the ICM magnetic field. The geometry and intensity of the magnetic field can determine the evolution of the stripped ISM \citep[e.g.][]{Gronke_2020,Sparre_2020,Ge_2021}. However, inferring its intensity is not trivial. In the followings we present different estimates aimed to map the magnetic field strength in the galaxy. \\

We evaluated the magnetic field at the stagnation point (i.e., the point in front of the galaxy where the local wind velocity is zero) with respect to the ICM to begin with. The draping framework \citep[][]{Pfrommer_2010,Sparre_2020} predicts, in case of a super-alfvenic motion (condition which is always satisfied by a super-sonic motion in a high-$\beta$ plasma such as the ICM), the formation of a strong magnetic field in front of the galaxy due to the accretion of a magnetic envelope composed of magnetised ICM. The magnetic field at the stagnation point would eventually reach a maximum value, $B_\text{max}$, of:
\begin{equation}
    B_{\text{max}}=\sqrt{8\pi\tilde{\alpha}\rho V_p^2}
    \label{b-stag}
\end{equation}
where $\tilde{\alpha}=2$ is a geometrical factor \citep[see][for further details]{Dursi_2008}, $\rho$ is the wind (i.e. ICM) density and $V_p$ is the post-shock velocity. In the JW100 condition in Abell 2626, where the ICM density at the clustercentric projected radius is $\rho=5.8\times10^{-27}$ g cm$^{-3}$ \citep[][]{Ignesti_2018}, Equation \ref{b-stag} yields $B_{\text{max}}\simeq5\times (V_p/100 \text{km/s})\text{ $\mu$G}$. Therefore for $V_p$ of the order of 2000 km s$^{-1}$ \citep[][]{Poggianti_2019} the stagnantation magnetic field could potentially reach extreme values ($\sim100$ $\mu$G). However, we note that such estimate is based on the ICM density estimated at the projected clustercentric distance of JW100, thus it represents an upper limit for the actual ICM density surrounding JW100. The draping magnetic field should decline in intensity with the distance from the contact surface \citep[][]{Sparre_2020}, thus we could expect lower values in the disk and, even further, in the tail.\\

We exploited two different methods to estimate the magnetic field within the disk. 
\begin{itemize}
\item We calculated the equipartition magnetic field  following the revised formulation proposed in \citet[][Equation 3]{Beck_2005}. We used the non-thermal flux density measured at 1.4 GHz and corrected for redshift \citep[see][]{Govoni-Feretti_2004} to evaluate the synchrotron intensity. The spectral flattening at low frequencies does not permit us to reliably constrain  the injection index, which is a direct proxy of the CRe energy distribution, thus we assumed a typical spectral index $\alpha=-0.7$. The CRe pathlength is also unknown, thus we tested two extreme values of 1 and 2 kpc \citep[][]{Krause_2018}. Under these assumptions, the resulting equipartition field lies between 11 and 13 $\mu$G. As a caveat, we note that the equipartition estimate is based on the assumption of the equilibrium of the radio source. This hypothesis may not be true in the current state of JW100 because the external pressure provided by the RPS could potentially affect the intensity of the magnetic field;

\item The theory of small-scale dynamo amplification predicts that the turbulence released by SN \citep[e.g.,][for a recent study]{Bacchini_2020} can locally amplify the magnetic field, thus inducing a connection between the latter and the SFR in the form of $B\propto SFR^{0.3}$ \citep[e.g.,][]{Gressel_2008,Schleicher_2013}. \citet[][]{Tabatabaei_2017} tested this hypothesis on the KINGFISHER sample \citep[][]{Kennicutt_2011} and provided an empirical relation:
    \begin{equation}
        \log(B)=(0.34\pm0.04)\log(\text{SFR})+(1.11\pm0.02)
    \label{b-sfr}    
    \end{equation}
For JW100, \citet[][]{Vulcani2018}  reports a SFR within the disk of 2.6$\pm0.5$ that yields $B=17.8\pm1.6$ $\mu$G. It is worth noticing that in the current condition of JW100 other source of turbulence related to the RPS should be available, such as the propagation of the bow-shock, which could further contribute to the magnetic field amplification \citep[e.g.,][]{Iapichino_2012};

\end{itemize}
To summarize, these procedures constrain the average magnetic field between 11 and 18 $\mu$G within the disk, which is in agreement with previous results for spiral galaxies \citep[e.g.,][]{Beck_2000,Tabatabaei_2017} but slightly higher than what is reported in \citet[][]{Muller_2021} for the jellyfish galaxy JO206 ($6.5-7.8$ $\mu$G). This could either be due to an intrinsic difference between the two galaxies (the stellar mass of JO206 is $9.1\times 10^{10}$ $M_\odot$ which is lower than JW100) or due to the uncertainties of the methods exploited to estimate the magnetic field.\\

\label{cool-l}
The morphology of the tail and the local ISM properties are complex and mostly unknown, which is contrary to the situation with the disk, hence the methods described above can not be reliably applied. Therefore, we estimated the magnetic field intensity by evaluating the cooling length. If the radio plasma is stripped from the disk, then cooling length (i.e., the radio tail truncation scale) roughly corresponds to $D\simeq t_{r}\times v_{\text{pl}}$, where $v_{\text{pl}}$ is the velocity of the relativistic plasma and $t_{R}$ is the radiative time of the CRe. Outside the stellar disk the CRe cooling should be mostly dominated by synchrotron and Inverse Compton losses due to the cosmic microwave background (CMB).   We assume that adiabatic losses are negligible because the stripped tails usually appear to be in quasi-equilibrium with the ICM \citep[e.g.,][]{Sun2010, Zhang2013,Campitiello_2021}. 
Therefore the radiative time-scale, $t_r$, can be expressed as: 
\begin{equation}
t_r\simeq3.2\times 10^{10}\frac{B^{1/2}}{B^2+B_{\text{CMB}}^2} \frac{1}{\sqrt{\nu (1+z)}}\text{ yr}
\label{cool_RW}
\end{equation}
where $B_{\text{CMB}}=3.25(1+z)^2$ $\mu$G is the equivalent CMB magnetic field and $\nu$ in units of MHz is the cut-off frequency \citep[e.g.,][]{Miley_1980}. Therefore, within this simple framework, the cooling length at a given frequency can be predicted based on the combination of the average magnetic field and the velocity of the CRe. Accordingly, the observed truncation scale could be used to jointly constrain these parameters. Due to the current resolution of our images we can not reliably measure the cooling length of the tail at 0.144 and 1.4 GHz (e.g., by fitting the surface brightness profile with an exponential model $I(r)\propto exp(r/r_c)$), as well as reliably detect it at 3.2 and 5.5 GHz. Therefore, we ultimately resolved to constrain the magnetic field by using the projected lengths observed at 0.144 and 1.4 GHz.\\

The $v_{\text{pl}}$ is unknown, therefore we propose two possible regimes:
\begin{itemize}
    \item Cloud velocity: the truncated radio disk, which has the same extension as the H$\alpha$ emission, suggests that the relativistic plasma follows the same dynamics of the stripped ISM clouds, which have a  characteristic velocity $v_{\text{pl}}=100-500$ km s$^{-1}$  \citep[e.g.,][]{Tonnesen_2021};
    \item Post-shock velocity: \citet[][]{Muller_2021} suggested that the CRe move along the ordered magnetic field (re-)accelerated by the bow-shock. The actual velocity of JW100 is unknown as well, but the estimate provided in \citet[][$\gtrsim$2200 km s$^{-1}$]{Poggianti_2019} entails a post shock velocity of $\sim$2000 km s$^{-1}$.
    
\end{itemize}
 Currently the lack of information on the polarization of the radio emission, hence on the geometry of the magnetic field, does not allow us test the latter regime. Moreover, the spectral steepening observed both in the disk and the tail (Figure \ref{spectra}) might suggest that the role of shock re-acceleration (i.e., spectral flattening at high frequencies) is negligible, hence that the contribute of the CRe moving at the post-shock velocity might not be relevant. Therefore, on the basis of the morphological similarities between the H$\alpha$ and the radio emission in the disk (which we further discuss in Section \ref{morph}), we favor the first scenario. \\
\begin{figure}
    \centering
    {\bf 0.144 GHz}
    \includegraphics[width=\linewidth]{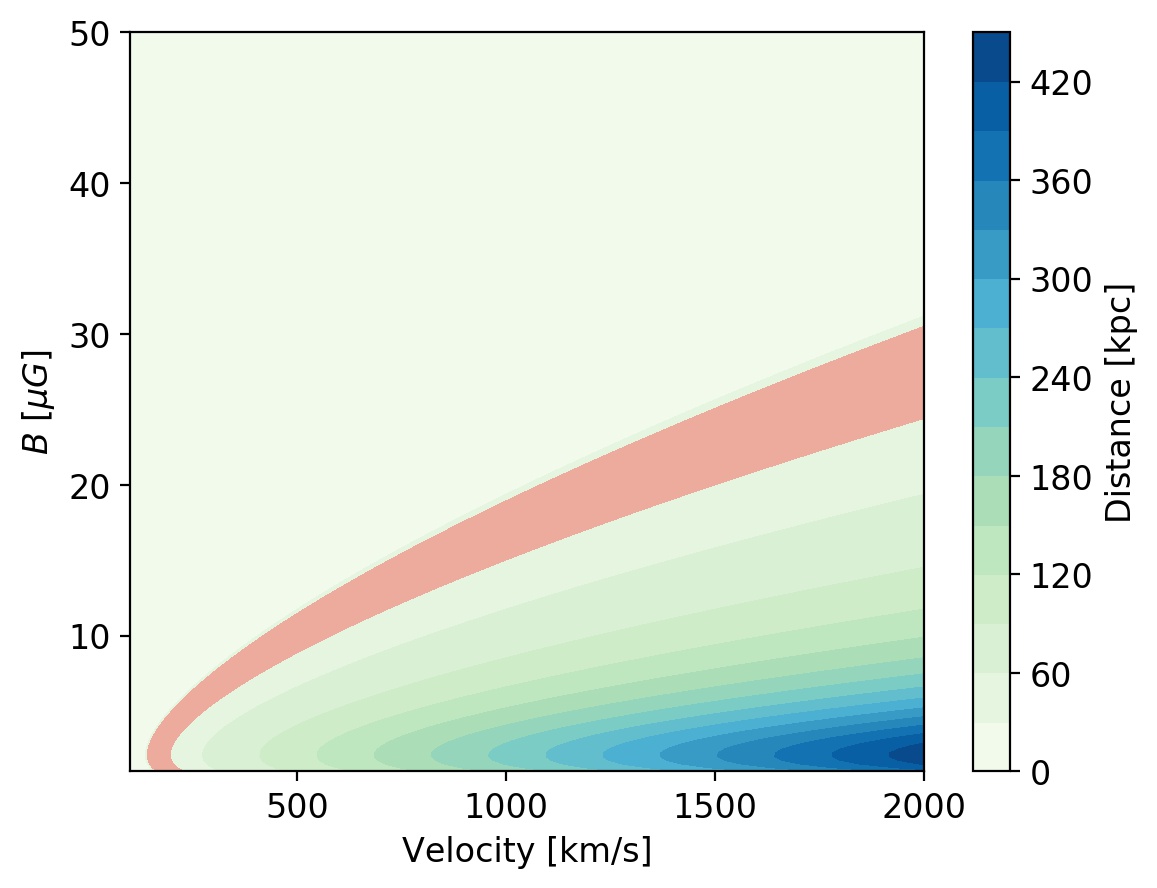}
    {\bf 1.4 GHz}
    \includegraphics[width=\linewidth]{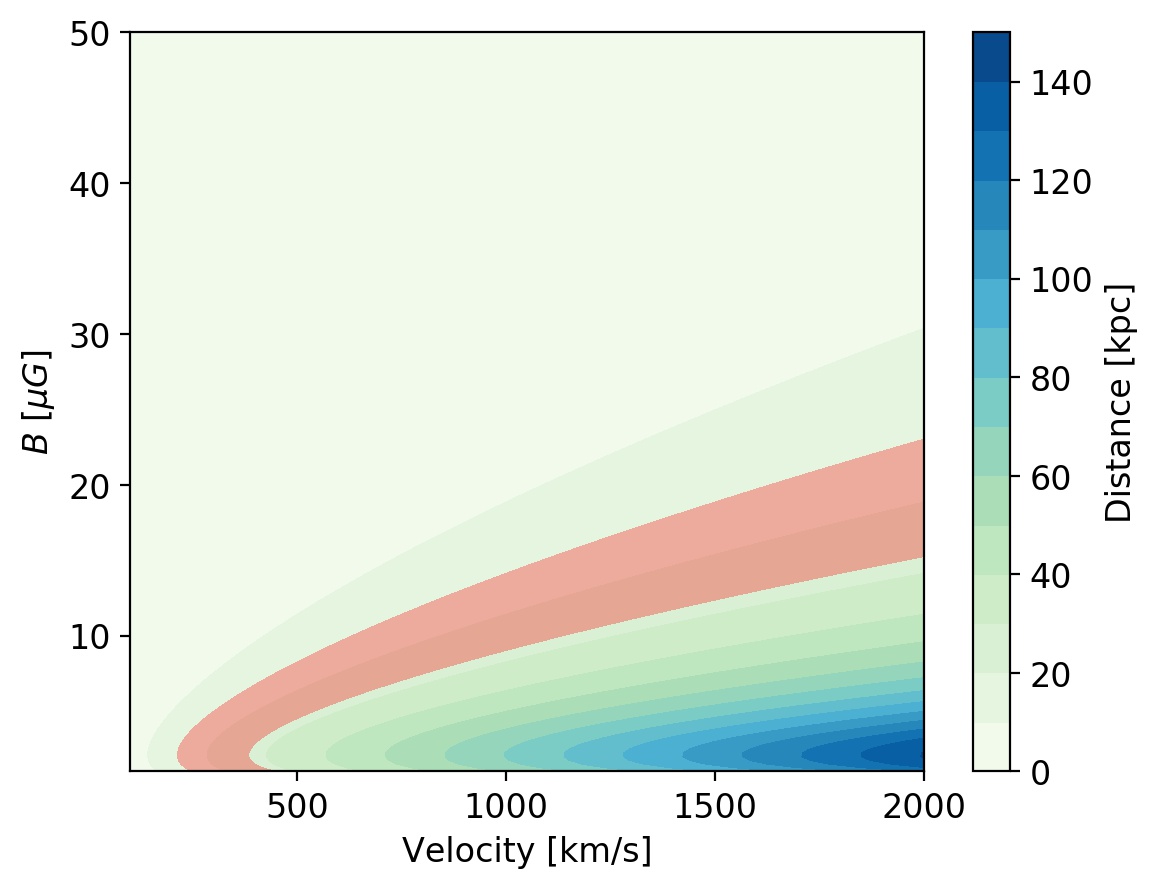}
    \caption{Cooling length parameter space for the radio tail observed at 0.144 (top) and 1.4 GHz (bottom) The red regions correspond to the observed projected lengths, here reported with a 6 kpc uncertainty which is due to the current resolution of 12$''$.}
    \label{col-length}
\end{figure}

This estimate has a series of caveats. To begin with, we are using the projected length of the tail, i.e. the lower limit of its actual extension, that is ultimately an observation-dependent quantity. Therefore, we can estimate only an upper limit for the magnetic field. Furthermore, we neglected any possible contribution to the CRe energy from the turbulent re-acceleration \citep[e.g.,][]{Brunetti-Lazarian_2016} within the tail, which could extend their life-time, or any injection of fresh CRe from the SN in the tail (i.e. we are assuming that the CRe are accelerated only within the disk). The former assumption could be supported by the fact that re-acceleration has been claimed to be negligible for high-mass galaxies \citep[][]{Roberts_2021a}, whereas the latter could be viable if the radio emission of the stripped, relativistic plasma accelerated in the disk overcomes the non-thermal emission associated with the SFR in the tail. Keeping in mind these limitations, by assuming  $v_{\text{pl}}=100-500$ km s$^{-1}$, we can match the projected tail lengths observed at 0.144 and 1.4 GHz which are 37 and 21 kpc respectively, for $B\leq$10 $\mu$G, as shown in Figure \ref{col-length}.\\

Interestingly, we note that, within this framework, by assuming (a) an uniform magnetic field and (b) that the CRe with different energy travel along the tail at the same velocity, the cooling length should change with frequency as:
\begin{equation}
    \frac{D_1}{D_2}=\frac{t_{\text{r,1}}}{t_{\text{r,2}}}\simeq\sqrt{\frac{\nu_2}{\nu_1}}
    \label{criterio-i}
\end{equation}
The current observations do not resolve enough of the tail of JW100 to test this, thus we refer to deeper, high-resolution observations of asymmetrical radio tails. Confirming (or not) the predictions of Equation \ref{criterio-i} would probe the CRe dynamics within the tail of jellyfish galaxies.

\section{Discussion}
The multi-frequency study presented in this work allows us to infer the characteristics of the non-thermal radio emission of JW100, hence providing us important insights into the properties of the CRe and the magnetic field. In the following, we use these new pieces of information, jointly with the results of the previous studies of this galaxy, to investigate a series of crucial questions, i.e. what is the origin of the CRe, what role the RPS can play in their evolution, and what can we learn about the evolution of this peculiar galaxy thanks to the properties of its non-thermal components. 

\subsection{Origin and evolution of the radio emitting electrons}
\label{sfr_disc}
\begin{figure}
    \centering
    \includegraphics[width=\linewidth]{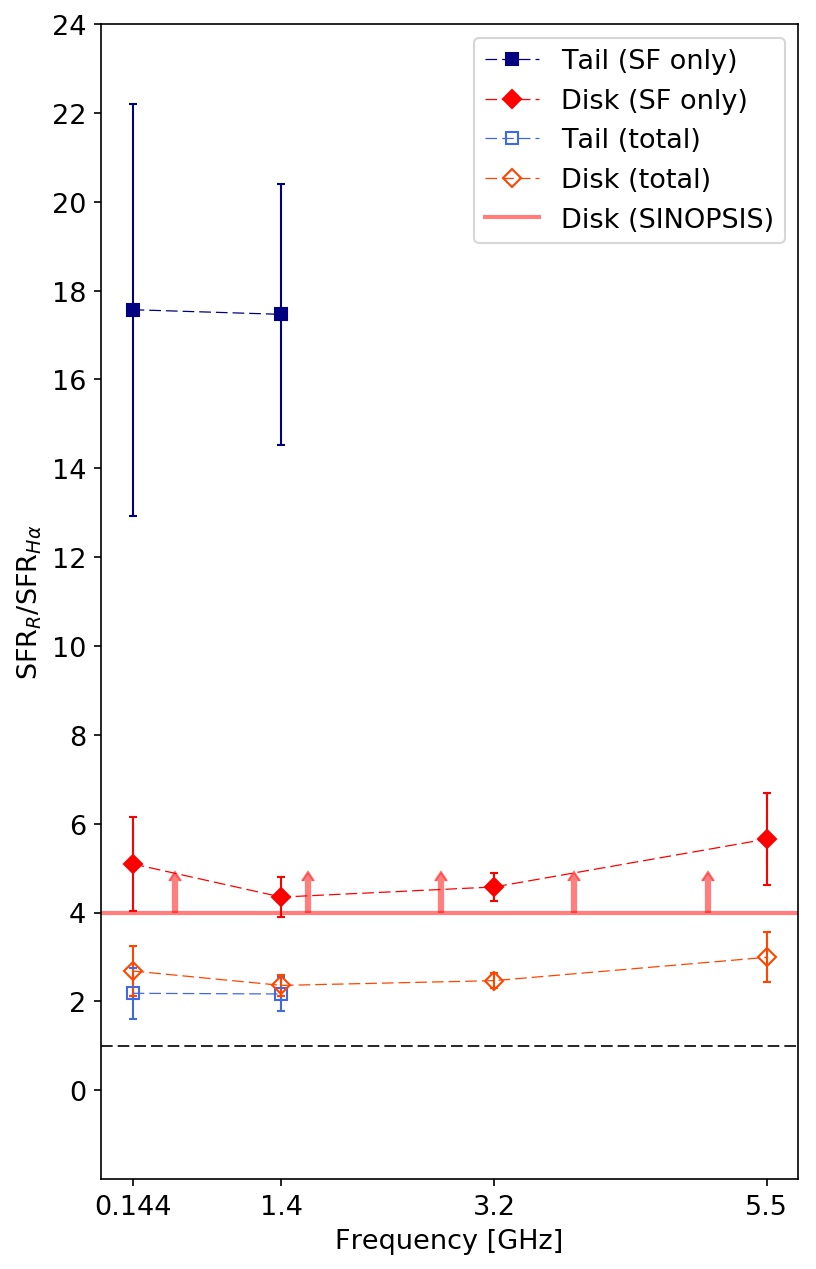}
    \caption{Ratio between SFR$_R$ and SFR$_{H\alpha}$ for disk (red) and tail (blue) at the different frequencies. For reference we report also the ratios computed by using the total H$\alpha$ emission (empty markers). The horizontal red line points out the median SFR ratio within the stellar disk derived from the SINOPSIS analysis (which is actually a lower limit, see Section \ref{sfr_disc}), whereas the black dashed line indicates the level SFR$_R$/SFR$_{H\alpha}=1$.}
    \label{sfr}
\end{figure}

In this Section we investigate the CRe origin by exploring their connection with the star formation history of the galaxy. Both radio and H$\alpha$ fluxes are SFR diagnostics \citep[e.g.,][for a review]{KennicuttEvans2012}, thus they can be used to compare the expected SFR in disk and tail. We estimated the radio-based SFR, SFR$_R$, at each frequency by adopting the conversion presented in \citet[][]{Garn_2009}:
\begin{equation}
    \text{SFR}_R=\frac{0.066}{1+z}\left(\frac{D_L(z)}{\text{Mpc}} \right)^2 \left[\frac{(1+z)\nu}{\text{1.4 GHz}} \right]^{-\alpha_\nu^{1.4}} \left(\frac{S_{\text{nt}}}{\text{Jy}} \right) \text{ $M_\odot$ yr$^{-1}$}
\end{equation}
where $D_L(z)$ is the luminosity distance at the redshift $z$, $\nu$ is the frequency, and $\alpha_\nu^{1.4}$ is the spectral index between $\nu$ and 1.4 GHz. This relation is an extension of the relation presented in \citet[][]{Bell_2003} and it is based on the \citet[][]{Chabrier2003} IMF. For the H$\alpha$-based estimate, SFR$_{H\alpha}$, we adopted the conversion presented in \citet[][]{Kennicutt1998a} accordingly converted to the \citet[][]{Chabrier2003} IMF:
\begin{equation}
    \text{SFR}_{H\alpha}=4.6\times10^{-42}\left(\frac{L_{H\alpha}}{\text{erg s$^{-1}$}} \right) \text{ $M_\odot$ yr$^{-1}$}
\end{equation}
where $L_{H\alpha}$ is the H$\alpha$ luminosity. We evaluated the SFR by using the respective fluxes measured from images with a matching resolution of 12$''$ and, in the case of the H$\alpha$, composed by selecting only those regions classified as star forming according to the [O I] BPT diagrams. For reference, the SFR ratios were evaluated also by using the total H$\alpha$ emission (see Figure \ref{opto}) accordingly smoothed to 12$''\times12''$. The results are presented in Figure \ref{sfr}.

We found that, in the disc, the SFR estimated from the radio is 5 times larger than the SFR estimated from the H$\alpha$ emission. In the tail, this discrepancy is even larger, a factor of 18. A similar discrepancy in the radio and H$\alpha$ signal between disks and tails has been observed also in the galaxies in the Coma cluster \citep[][Figure 6]{Chen_2020}. When the total H$\alpha$ emission is included, the excesses decrease to a factor $\times2.5$ for both the disk and the tail. Nevertheless, the excess of the radio emission with respect to the observed H$\alpha$ emission persists.\\

The excess in the tail could be explained as a further evidence that the relativistic plasma we observe in that region had been stripped from the disk, thus is not related to the local star formation traced by the H$\alpha$ emission. The discrepancy in the disk could instead be interpreted as evidence of a higher SFR in the past or, in other words, of recent quenching in the galaxy. Indeed, the total radio emission can trace the average SFR within the typical radiative time of the CRe (i.e. few $10^7$ yrs), whereas the H$\alpha$ emission traces the most recent phase ($<10^7$ yrs) of star formation.\\

\begin{figure}
    \centering
    \includegraphics[width=\linewidth]{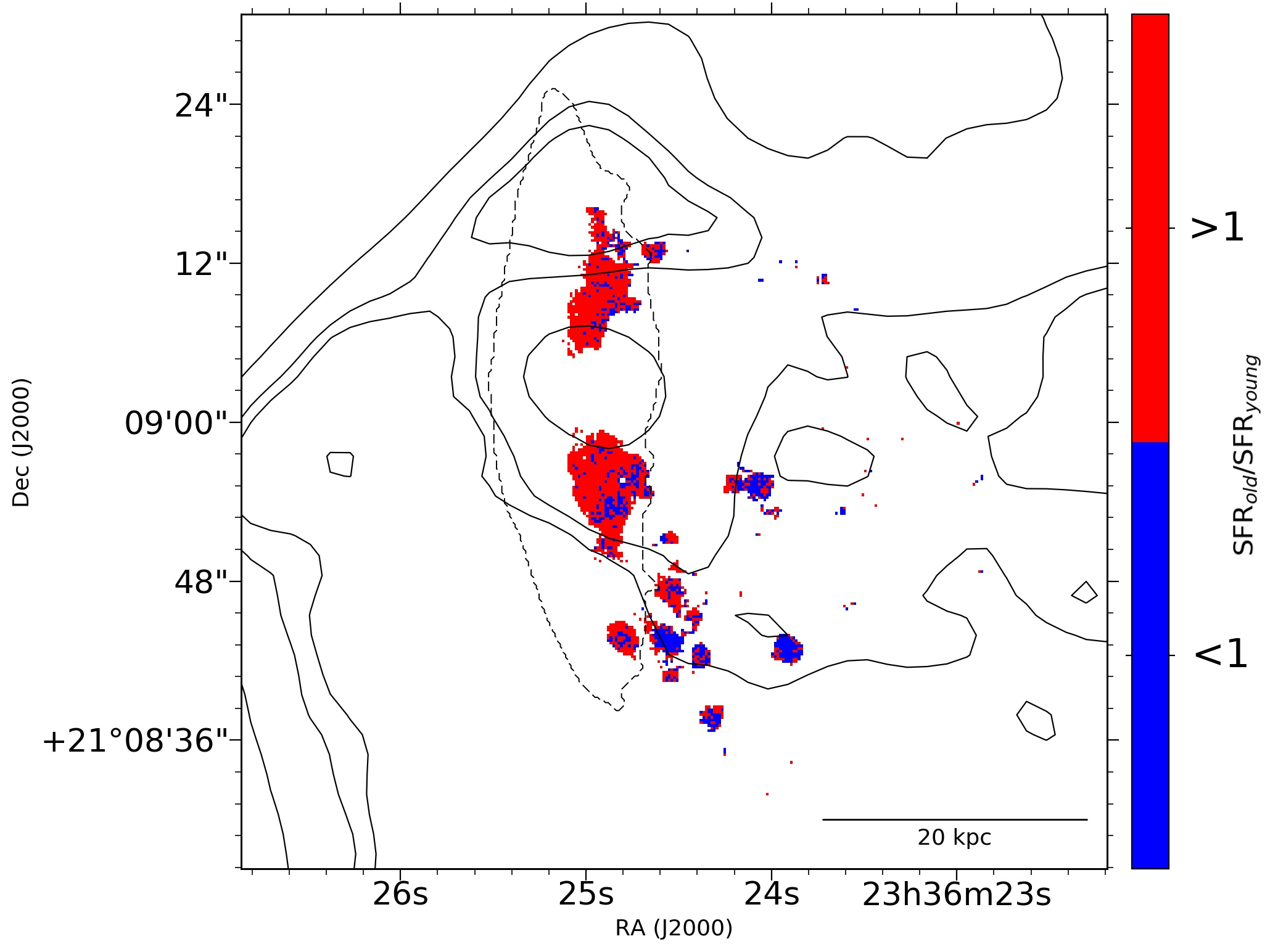}
    \caption{Ratio between the SFR within $5.7\times10^7$ and $2\times10^6$ yrs derived with SINOPSIS. We report the values lower than 1 in blue, and above  1 in red. For reference, we report also the contours of the 0.144 GHz radio emission shown in Figure \ref{canvas} (continuous) and the stellar continuum (dashed).}
    \label{SINOPSIS}
\end{figure}

In order to test this scenario we estimated the SFR in two different epochs by using the SINOPSIS analysis \citep[][]{Fritz2017}. SINOPSIS is a spectrophotometric code  that searches the combination of Single Stellar Population (SSP) model spectra that best fits the equivalent widths of the main lines in absorption and emission and the continuum at various wavelengths, minimizing the $\chi$ = 1 using an adaptive simulated annealing algorithm \citep{Fritz2007, Fritz2011}. The star formation history is let free with no analytic priors.
SINOPSIS uses a \cite{Chabrier2003} IMF with stellar masses in the 0.1-100 M$_\odot$ limits, and they cover metallicity values from Z = 0.0001 to 0.04. The metallicity of the best fit models is constant and homogeneous (i.e. all the SSPs have the same metallicity independently of age). The best fit model is searched using SSP models at three different metallicity values (sub-solar, solar and super-solar). Dust extinction is accounted for by adopting the Galaxy extinction curve \citep{Cardelli1989}. SINOPSIS uses the latest SSP models from S. Charlot \& G. Bruzual (in prep.) based on stellar evolutionary tracks from \cite{Bressan2012} and stellar atmosphere spectra from a compilation of different authors, depending on the wavelength range, stellar luminosity, and effective temperature. SINOPSIS also includes the nebular emission lines for the young (i.e., age $<$2$\times 10^7$ yr) SSPs computed with the Cloudy code \citep{Ferland2013}. Among other quantities, the code provides, for each MUSE spaxel, the average SFR in twelve age bins. These bins can be combined into larger bins in such a way that the differences between the spectral characteristics of the stellar populations are maximal \citep{Fritz2017}. For the aim of our analysis we estimated the time-averaged SFR within two intervals, $<5.7\times10^7$ (SFR$_\text{old}$) and $<2\times10^6$ yrs (SFR$_\text{young}$), which roughly correspond to the time-scales traced by the radio and H$\alpha$ emissions, and we computed the pixel-by-pixel ratio between them in the star-forming spaxels (Figure \ref{halpha}), which we report in Figure \ref{SINOPSIS}.\\

Our analysis revealed that the SFR in the disk was generally higher in the past (SFR$_\text{old}/$SFR$_\text{young}>$1). Specifically, in the star forming regions within the stellar disk we measure a median value of 4 that is close to the observed radio excess observed in the same regions of the disk (Figure \ref{sfr}). We note that this value represents a lower limit for the actual ratio. Due to our selection of the star forming regions, we are neglecting the past SFR of a large portion of the disk that has stopped forming stars only recently, hence it is not traced by the $H\alpha$ emission. Moreover, by using the [O I] classification, we are more conservative in the selection of the star forming regions than by using the other BPT diagrams. We also note the time intervals we are considering are relatively young, and therefore prone to degeneracies in age and dust attenuation in the SINOPSIS analysis. \\

On the basis of our findings, we suggest that the radio emission in the disk is actually dominated by a radiatively ``old" plasma with a radiative age of few $10^7$ yrs. Therefore, the spectral steepening observed in the disk is the result of the combination of the radio emissions produced by the old, steep-spectrum plasma accelerated during the past star-forming phase, and the young, flat spectrum plasma related to the ongoing star formation.  Accordingly, the steeper spectrum we observe in the tail could be the consequence of a combination of two effects. On the one hand, the radio plasma naturally loses its energy while it travels to the tail, thus resulting in a steeper-spectrum radio emission (e.g., for a stripping velocity of $100-500$ km s$^{-1}$, the time required to cover 10 kpc would be of the order of few $10^7$ yrs, which corresponds to the radiative time of the CRe emitting at GHz frequencies). On the other hand, by moving in a region with a lower magnetic field ($<10$ $\mu$G, see Section \ref{magnetic}), the CRe emit at lower frequencies and thus the non-thermal spectrum gets shifted to lower frequencies (i.e., these observations sampled the steeper part of the spectrum). We also note that the SINOPSIS analysis revealed that the blobs outside the disk are dominated by the more recent star formation (i.e. a ratio between the SFR within $5.7\times10^7$ and $2\times10^6$ yrs lower than 1 in Figure \ref{SINOPSIS}). This indicates that these star forming blobs are recent ($<2\times10^6$ yrs), hence that the stripped ISM can survive, cool down and form new stars also outside the disk \citep[see][]{Vulcani2018,Poggianti_2019,Moretti_2020}. As we discuss in Section \ref{morph}, the local magnetic field can play a role in this by protecting the cold ISM from the hot, hostile ICM.\\

On the basis of the SINOPSIS analysis we propose that radio excess in the disk could be explained by a fast quenching of the galaxy, where the SFR decreased by a factor 4 within the last few 10$^7$ yrs. A similar scenario of star formation quenching, or lower star formation efficiency, within the disk of JW100 has been proposed also in \citet[][]{Moretti_2020}. They observed that the regions characterized by a lower star formation efficiency are located at the center in the proximity of the galactic bulge. The presence of a stellar bulge, as well as the RPS, has been observed to be systematically able to suppress the SFR \citep[e.g.,][]{Gensior_2020}, hence we suggest that the SFR$_R$-SFR$_{H\alpha}$ discrepancy could be the natural outcome of this.  \\

However, we can not exclude that the radio excess could depend also on other factors. The compression of the ISM, i.e. of the magnetic field, by the ram pressure would result in an enhancement of the radio emission \citep[e.g.,][]{Markevitch_2005}, thus in a higher estimate of SFR$_R$. This scenario was proposed to explain the general excess of radio emission observed in cluster late-type galaxies \citep[e.g.,][]{Gavazzi_1999}. Another possible source of additional radio emission could be the acceleration of galactic CRe as result of the interactions with the ICM winds \citep[][]{Murphy2009} or an additional component of radio emission associated with the magnetized drape. Alternatively, due to the low resolution we can not completely exclude that there is a contribution of the AGN radio emission in the disk region. 

 \subsection{Insights into the ISM-ICM interplay}
\label{morph}
\begin{figure*}
    \centering
    \includegraphics[width=.7\linewidth]{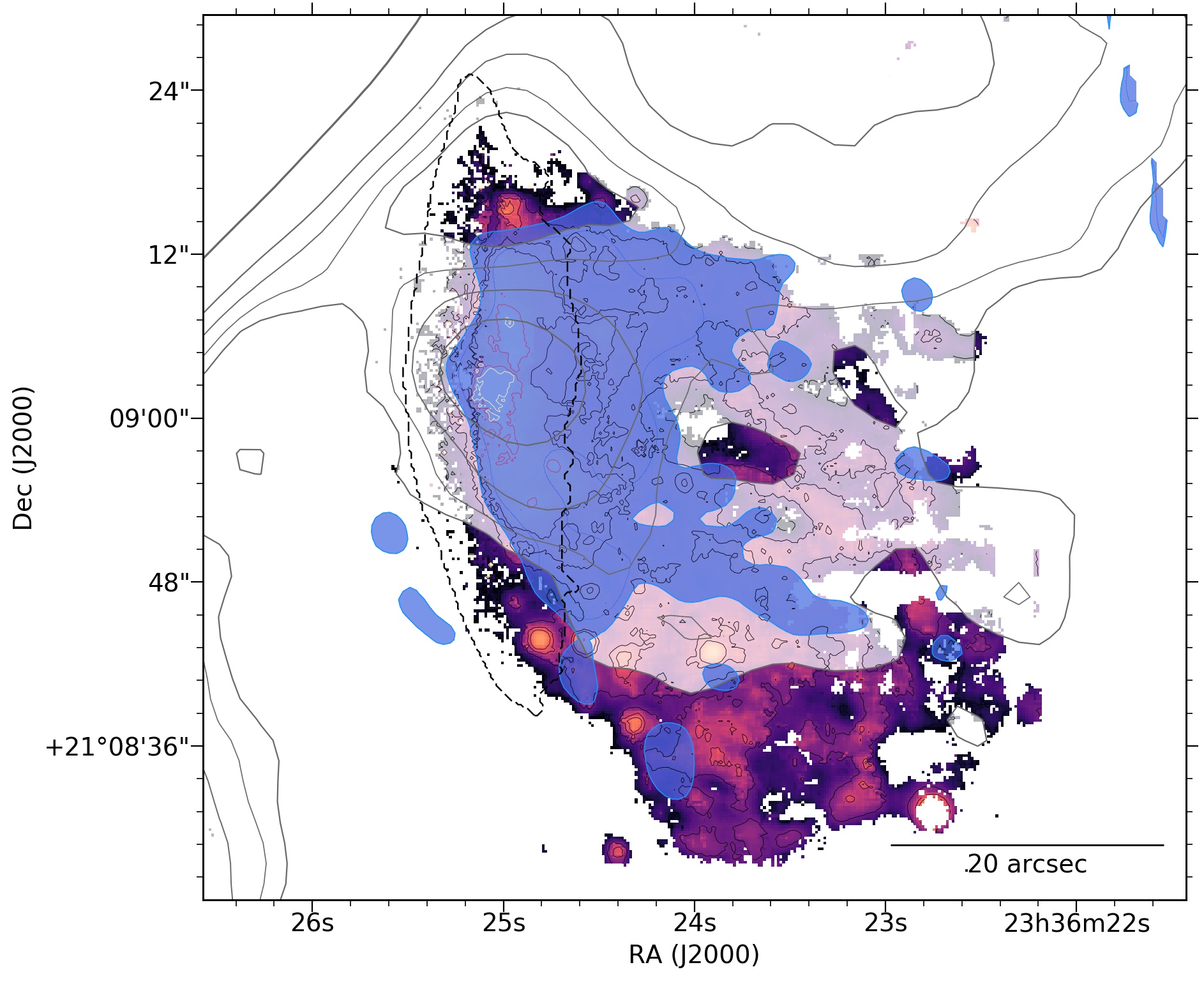}
    \caption{Multi-wavelength image of JW100: MUSE stellar continuum (dashed black line) and H$\alpha$ (color map, resolution 0.2$''$) with the contours of the 0.144 GHz emission from Figure \ref{canvas} (white-filled grey contours) and the {\it Chandra} X-ray image in the 0.5-2.0 keV  smoothed with a 1.5$''$ Gaussian (blue, filled).}
    \label{x}
\end{figure*}

By combining radio, H$\alpha$, and X-ray images of JW100 (Figure \ref{x}) we can explore the connection between different ISM components, namely the relativistic plasma, the warm ISM, and the hot galactic plasma likely produced by the interplay between ISM and ICM \citep[e.g.,][]{Sun2010,Poggianti_2019,Campitiello_2021,Sun_2021}. We observe that the disk truncation is a common characteristic among radio, H$\alpha$, and X-ray emissions, thus hinting that (1) ram pressure stripping is equally relevant for the three corresponding components or (2) both the radio and the X-ray extraplanar emissions are somehow consequence of the warm ISM being stripped. 
The first hypothesis could be consistent with a scenario where the relativistic plasma is stripped along the ISM clouds (e.g. in form of magnetic field and CRe traveling within the stripped ISM clouds). The second one, instead, may be related to the ICM draping \citep[][]{Dursi_2008, Pfrommer_2010}. According to the draping scenario, the magnetized layer is composed of hot ICM. As a consequence of the density fluctuations induced by the passage of the bow-shock, the magnetized layer can cool down and results in extended X-ray emission \citep[][]{Sparre_2020}. Therefore, the spatial correlation between radio and X-ray emission would be expected because these emissions are produced within the same hot, magnetized  layer. Interestingly, the lack of flat spectrum emission outside the star-forming regions may suggest that the CRe (re-)accelerated by the bow-shock marginally contribute to the radio emission, thus questioning to what degree the bow shock may have actually affected the CRe. We refer to future polarimetric studies, which are beyond the scope of the present work, to further test this scenario by probing the presence of the signature ordered field (i.e., a high degree of extended polarized emission aligned along the tail) produced by the ICM draping.\\

Outside the disk, the spatial correlation between these phases seems to decrease, in the sense that the radio and X-ray emission seem to elongate mostly toward the west\footnote{Albeit the 0.144 GHz radio emission seems to be more extended toward west than the H$\alpha$, we note that this is likely due to the different resolutions and the fact that the western edge of the radio tail falls outside the MUSE CCD.}, whereas the H$\alpha$ emission is more extended toward the south-west. The discrepancy between the different emissions could be due to sensitivity issue, or due to the different time-scales of the emissions. Specifically, the lack of radio emission in the southern part of the tail could indicate that this region is older than a few radiative-times of the 0.144 GHz emission (i.e. few $10^8$ yrs), thus without injection of fresh CRe the radio emission is no longer visible in our images. Noticeably, the lack of X-ray emission in the farthest part of the tail might suggest that there the interaction between ISM and ICM, and the cooling of the latter, is dominated by mixing instead of shock-induced radiative cooling \citep[e.g.,][]{Gronke_2018,Kanjilal_2020}. However, investigating the physics of mixing is extremely complex and beyond the aim of this work. Interestingly, the magnetic field tentatively constrained by the cooling length ($\leq 10$ $\mu$G, see Section \ref{magnetic}) would potentially be strong enough to preserve the stripped ISM from the interactions with the ICM, in the form of conduction or hydrodynamical instabilities \citep[e.g.,][]{Berlok2019,Cottle_2020}, thus allowing the stripped clouds to survive outside the disk and, potentially, form new stars \citep[e.g.,][]{Sparre_2020,Ge_2021}. 

\subsection{On the AGN radio emission}
\label{agn}
\begin{figure}
    \centering
    \includegraphics[width=\linewidth]{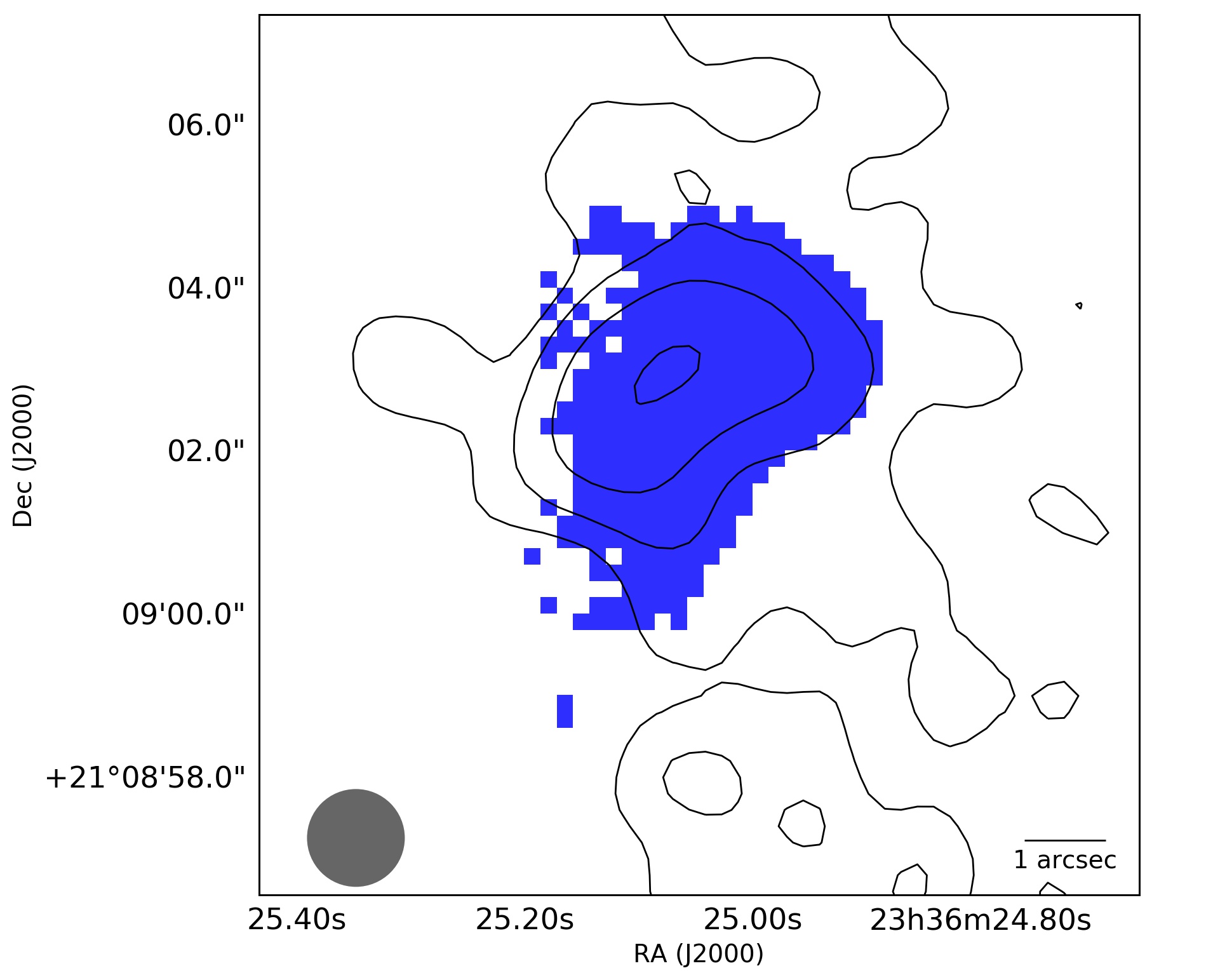}
    \caption{ A zoom into the AGN region selected according to [O I] classification \citep[][]{Poggianti2019} with the 3,  6, 12, 24$\sigma$ contours from the VLA observations at 1.4 GHz presented in \citet[][]{Gitti_2013b} (resolution 1.3$\times$1.3 arcsec as reported in the left-bottom corner, and $1\sigma=$12.9 $\mu$Jy beam$^{-1}$) .}
    \label{high-res}
\end{figure}
JW100 hosts a central AGN (Seyfert2) with evidence of optical \citep[][]{Poggianti_2019,Poggianti2019}, radio \citep[][]{Gitti_2013b,Ignesti_2017} and X-ray \citep[][]{Wong_2008, Poggianti_2019} emissions. The AGN is characterized also by the presence of an ionized gas outflow \citep[][]{Radovich2019}. In agreement with the high-resolution images presented in \citet[][]{Gitti_2013b,Ignesti_2017}, we detect a peak in surface brightness coinciding with its position. We notice that the morphology of the radio emission derived from the high-resolution VLA image of Abell 2626 presented in \citet[][]{Gitti_2013b} coincides remarkably well with the optical emission due to the AGN, traced by the [OI] BPT diagram  \citep[][]{Poggianti2019}. Kiloparsec-scale radio emission in Seyfert galaxies is common \citep[][]{Gallimore_2006}, and it is explained by the `frustrated jet model'. This scenario predicts that the kinetic energy of the relativistic jets can be transferred to the ISM, possibly triggering a gas outflow from the inner region of the galaxy \citep[we refer to ][ for a more detailed formulation]{Bicknell_1997}. Taking into accounts the caveats reported in Section 3.2, the AGN non-thermal spectrum allows us to tentatively explore the connection between the origin of the radio emission and the ionized gas outflow observed in JW100.\\ 

\begin{figure}
    \centering
    \includegraphics[width=\linewidth]{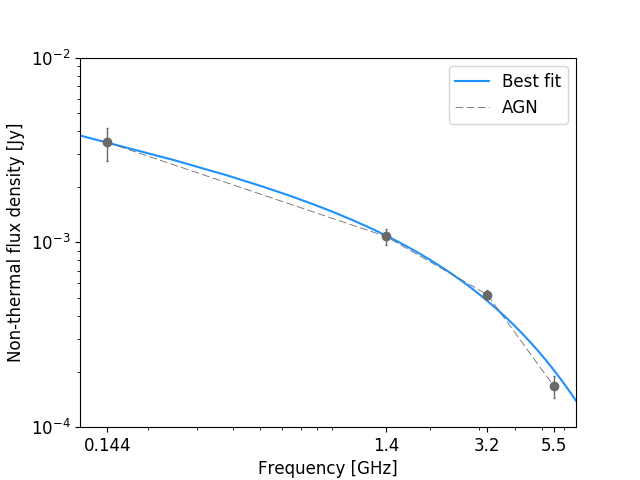}
    \caption{AGN non-thermal spectrum from Figure \ref{spectra} and the best-fitting exponential cut-off model.}
    \label{fit}
\end{figure}

Our new multi-frequency analysis revealed that the AGN is characterized by a steep spectrum above 1.4 GHz (Figure \ref{spectra}) that could be an indication of a radiatively-old relativistic plasma. We tested this hypothesis by fitting the observed non-thermal spectrum with an exponential cut-off model, which is suited to fit the exponential decline produced by energy losses \citep[e.g.,][]{Eilek_1996}:
\begin{equation}
    S(\nu)\propto\nu^{\alpha}e^{-\frac{\nu}{\nu_{b}}}
\end{equation}
By fitting the observed spectrum we estimate a break frequency $\nu_{b}$=3.4$\pm$0.8 GHz (Figure \ref{fit}). For $B=11-18$ $\mu$G CRe emitting at this frequency have typical radiative ages of 0.7-1.3$\times10^7$ yrs (Equation \ref{cool_RW}), which is in agreement with the age of the ionized gas outflow estimated in \citet[][$0.8\times 10^7$ yrs]{Radovich2019}. Therefore we suggest that the ejection/acceleration of the radio plasma within the AGN region, possibly due to the presence of unresolved, frustrated radio jets, might be linked to the arcsec-scale ionized gas outflow observed by MUSE.

\section{Summary and Conclusions}
We presented a detailed analysis of the radio emission of the jellyfish galaxy JW100 based on an unprecedented assembly of radio observations. By combining LOFAR, MeerKAT, VLA, and VLA observations, as well as previous results obtained with MUSE and {\it Chandra}, we  investigated the properties and the origin of the radio-emitting, relativistic plasma. We studied the spectrum of the star-forming regions within the disk, finding that below 1.4 GHz the non-thermal spectrum shows signatures of the results of several energy loss mechanisms, whereas at higher frequencies it shows the characteristic steepening ($\alpha=-2.40\pm0.36$ between 3.2 and 5.5 GHz) produced by an a advanced radiative stage. By comparing our findings with the results of the SINOPSIS analysis we concluded that the non-thermal emission in the disk is dominated by the old CRe injected when the SFR was, at least, a factor $\times4$ higher than currently inferred from the H$\alpha$ emission. We also estimated the magnetic field in the galaxy using different methods. We observed a decrease of the magnetic field from the disk (11-18 $\mu$G) to the tail ($\leq10$ $\mu$G). Interestingly, the magnetic field in the tail would be strong enough to extend the life of stripped ISM in the ICM by protecting the cold gas from thermal conduction and hydrodynamical instabilites. Finally we investigated the synchrotron spectrum of the AGN finding that its radiative age is in agreement with the age of the ionized gas outflow, thus suggesting a connection between the acceleration of the relativistic plasma and the gas outflow.\\

The study presented here offers a series of new insights into the physics of ram pressure stripping provided by the non-thermal radio emission. The spatial correlation observed between the radio and X-ray emission might suggest a physical connection between these phases, which would be in agreement with the ICM draping scenario. Moreover, our multi-wavelength analysis proved that the non-thermal radio emission can be a powerful tool to probe the star formation history of a galaxy. On the basis of the excess of radio emission with respect to the current star formation, we could confirm the fast quenching of JW100. We further argue that a similar scenario could likely have taken place in the other high-mass jellyfish galaxies, thus the radio excess observed in ram pressure stripped galaxies could be interpreted as well as the result of fast quenching of the SFR.\\

More insights will come from future high-resolution observations, such as LOFAR observations with a 0.3$''$ resolution performed with the international stations, and studies carried out with the upcoming Square Kilometer Array, that will allow a more detailed comparison with the optical and X-ray observations. Furthermore, deep observations designed for polarimetric studies at higher frequencies are now crucial to address the open questions about the geometry of the magnetic field posed by the present work.

\section*{Acknowledgements}
We thank the anonymous Referee for their thoughtfully work that improved the presentation of the manuscript.  AI thanks I. Ruffa for the useful discussions. This work is the fruit of the collaboration between GASP and the LOFAR Survey Key Project team (``MoU: Exploring the low-frequency side of jellyfish galaxies with LOFAR", PI A. Ignesti, B. M. Poggianti, S. McGee). AI, BV, RP, MG acknowledge the Italian PRIN-Miur 2017 (PI A. Cimatti). MV acknowledges support by the Netherlands Foundation for Scientific Research (NWO) through VICI grant 639.043.511. IDR and RJvW acknowledge support from the ERC Starting Grant Cluster Web 804208. AB acknowledges support from the VIDI research
programme with project number 639.042.729, which is financed by the Netherlands Organisation for Scientific Research (NWO). SLM acknowledges support from the Science and Technology Facilities Council through grant number ST/N021702/1. JF acknowledges financial support from the UNAM- DGAPA-PAPIIT IN111620 grant, México. We acknowledge funding from the INAF main-stream funding programme (PI B. Vulcani). Based on observations collected at the European Organization for Astronomical Research in the Southern Hemisphere under ESO programme 196.B-0578. This project has received funding from the European Research Council (ERC) under the European Union's Horizon 2020 research and innovation programme (grant agreement No. 833824). LOFAR data products were provided by the LOFAR Surveys Key Science project (LSKSP; https://lofar-surveys.org/) and were derived from observations with the International LOFAR Telescope (ILT). LOFAR \citep[][]{vanHaarlem_2013} is the Low Frequency Array designed and constructed by ASTRON. It has observing, data processing, and data storage facilities in several countries, which are owned by various parties (each with their own funding sources), and which are collectively operated by the ILT foundation under a joint scientific policy. The efforts of the LSKSP have benefited from funding from the European Research Council, NOVA, NWO, CNRS-INSU, the SURF Co-operative, the UK Science and Technology Funding Council and the Jülich Supercomputing Centre. This paper makes use of the MeerKAT data (Project ID: SCI-20190418-JH-01). The MeerKAT telescope is operated by the South African Radio Astronomy Observatory, which is a facility of the National Research Foundation, an agency of the Department of Science and Innovation. AI thanks the Black Sabbath's music for providing the inspiration during the preparation of the draft.

\bibliography{bibliography}{}
\bibliographystyle{aasjournal}



\end{document}